\documentclass{emulateapj}

\usepackage{natbib}

\newcommand{\mpc}{\ensuremath{\, h^{-1}\,\mathrm{Mpc} }}
\newcommand{\kpc}{\, h^{-1}\,\mathrm{kpc} }
\newcommand{\kms}{\, \mathrm{km \; s^{-1}}}
\newcommand{\stwo}{\ensuremath{S_2}}
\newcommand{\dstwo}{\ensuremath{\Delta S_2}}
\newcommand{\ctwo}{\ensuremath{C_2}}
\newcommand{\dtwo}{\ensuremath{D_2}}
\newcommand{\stwohat}{\ensuremath{\hat{S}_2}}
\newcommand{\dtwohat}{\ensuremath{\hat{D}_2}}
\newcommand{\ctwohat}{\ensuremath{\hat{C}_2}}
\newcommand{\lya}{Ly$\alpha$}
\newcommand{\fth}{\ensuremath{F_{th}}}
\newcommand{\heii}{\ion{He}{2}}
\newcommand{\scien}[2]{#1  \times 10^{#2}} 
\newcommand{\be}{\begin{equation}}
\newcommand{\ee}{\end{equation}}
\newcommand{\bc}{\begin{center}}
\newcommand{\ec}{\end{center}}
\newcommand{\bfig}{\begin{figure}}
\newcommand{\efig}{\end{figure}}
\newcommand{\snr}{\ensuremath{\mathrm{S/N}}}
\newcommand{\fbar}{\ensuremath{\overline{F}}}
\newcommand{\fmean}{\ensuremath{\langle F \rangle}}
\newcommand{\lnl}{\ensuremath{-\ln \mathcal{L}} }

\slugcomment{Accepted by ApJ}
\shorttitle{Threshold Probability Functions and the Lya Forest}
\shortauthors{Lee \& Spergel}

\begin{document}
\title{Threshold Probability Functions and \\ Thermal Inhomogeneities in the 
Lyman-alpha Forest}
\author{Khee-Gan Lee and David N. Spergel}
\affil{Department of Astrophysical Science, Princeton University, Princeton, New Jersey 08544, USA}
\email{lee@astro.princeton.edu}
\email{dns@astro.princeton.edu}

\begin{abstract}
We introduce to astrophysics the threshold probability functions \stwo,\ctwo,
and \dtwo\ first derived by
\citet{torq+88}, which effectively
samples the flux probability distribution (PDF) of the \lya\ forest at different spatial scales. 
These statistics are tested on mock \lya\ forest spectra based on various toy models for 
\heii\ reionization, with homogeneous models with various temperature-density relations as well as
models with temperature inhomogeneities. These mock samples have systematics and noise added to
simulate the latest Sloan Digital 
Sky Survey Data Release 7 (SDSS DR7) data.
 We find that the flux PDF from SDSS DR7 can be used to constrain 
the temperature-density relation $\gamma$ (where $T \propto (1 + \Delta)^{\gamma-1}$) 
of the intergalactic medium (IGM) at $z=2.5$ to a precision of 
$\Delta \gamma = 0.2$ at $\sim 4\sigma$ confidence. The flux PDF is degenerate to temperature 
inhomogeneities in the IGM arising from \heii\ reionization,
 but we find \stwo\ can detect these inhomogeneities at $\sim 3 \sigma$,
with the assumption that the flux continuum of the \lya\ forest 
can be determined to 9\% accuracy,
approximately the error from current fitting methods. 
If the flux continuum can be determined to 3\% accuracy, then \stwo\ is capable of constraining
the characteristic scale of temperature inhomogeneities, with $\sim 4 \sigma$ differentiation 
between toy models with hot bubble radii of 50 \mpc\ and 25 \mpc.
\end{abstract}


\section{Introduction}

The absorption of radiation by the Lyman-$\alpha$ (\lya) resonance of 
neutral hydrogen along the
line-of-sight to high-redshift quasars,
commonly known as the \lya\ forest,  is an important probe of
large-scale structure at $z \gtrsim 2$
\citep[see, e.g.,][]{croft+98,mcd+00,croft+02,zald+03,mcd+05}. 
The utility of the \lya\ forest as a cosmological tool has been enabled by 
theoretical work on the intergalactic medium (IGM)
\citep[see, e.g.,][]{cen+94,mirald+96,croft+98,dave+99,theuns+98},
 which allowed the 
underlying dark matter field to be mapped from the \lya\ absorption. 

In recent years, increasing attention has been turned towards 
obtaining a deeper understanding on the detailed astrophysics of the IGM 
such as the ionizing ultraviolet background,  temperature field, 
metals, etc. The reionizations of \ion{H}{1} and \heii\ have been shown to 
play critical roles in regulating the properties of the IGM.

The reionization of \heii\ at $z \sim 3$ should leave an observable imprint on
the properties of the \lya\ forest.
 Recent theoretical developments have recognized
that \heii\ reionization must have occurred in an extended and inhomogeneous
fashion as the process was driven by rare and bright quasars \citep{lai+06,furl+oh08, mcquinn+09}.  
These spatial variations in \heii\ reionization history should modulate the equation of state
and entropy of the IGM.

Observations of the \lya\ forest have the potential to reveal details of this reionization
process.
The sources of $E > 54.4 \mathrm{eV}$ photons which drive \heii\ reionization
are believed to be rare bright quasars, but different model assumptions on, e.g., quasar
duty cycles and luminosity functions can dramatically change the history 
and morphology of \heii\ reionization \citep{mcquinn+09}, as well as the thermal
properties of the resulting IGM \citep{furl+oh08-2}.

Observational constraints on \heii\ reionization have arguably lagged behind the 
theoretical work. \citet{schaye+00} measured a peak at $z \approx 3$ 
in the temperature evolution of the IGM from the Doppler parameters of high-resolution
\lya\ forest spectra, that they claimed to be due to \heii\ reionization. 
While some authors \citep{theuns+02,bern+03,fg+08} argue that
the detection of the feature at $z \sim 3.2$ in the evolution of the effective \lya\ 
optical depth, $\tau_\mathrm{eff}$,  of the IGM provides further evidence
for the reionization transition, \citet{dall+09} failed to find this transition.
More recently, the Cosmic Origins Spectrograph on the 
Hubble Space Telescope is beginning to shed light on \heii\ reionization
through studies of the \heii\ \lya\ forest and associated \heii\ 
Gunn-Peterson troughs \citep[see, e.g., ][]{schull+10}.

Measurements of the flux probability distribution function \citep[PDF][]{jen+ost91} of high-resolution
\lya\ forest spectra \citep[e.g., ][]{mcd+00, lidz+06,kim+07} 
have constrained the equation of state
$\gamma$ of the IGM. In this paper, we define the equation of state
of the gas as a function only of the gas density:
\be \label{eq:eos}
T(\Delta) = \bar{T}\; \Delta^{\gamma-1},
\ee
where $T(\Delta)$ is the temperature as a function of density, $\bar{T}$ is the 
temperature at mean density, and $\Delta = \rho/\langle \rho \rangle$ is the density
contrast of the gas.
Different scenarios for \heii\ reionization make distinctive predictions for the redshift
evolution of $\gamma$.
Several authors \citep{becker+07, viel+09} claimed to have detected 
an inverted equation of state $\gamma < 1$, which is has been theorized to arise
 from the late reionization of voids \citep{furl+oh08-2}. 
\citet{lidz+09} attempted to measure spatial inhomogeneities in the thermal state of the IGM
by using a wavelet filter on high-resolution spectra, but had a null detection. 

Most observational attempts to place constraints on the IGM have been based on 
relatively small-numbers of
high-resolution ($R \equiv \lambda / \Delta \lambda \sim 10^4$) and high signal-to-noise
($\snr \sim 10^2$ per pixel) spectra. 
However, the largest single source of data on the \lya\ forest is arguably the 
Sloan Digital Sky Survey\footnote{\url{http://www.sdss.org}} \citep{york+00}, which
 includes $\sim 10^4$ quasars with useable \lya\ forest, 
albeit of moderate quality ($R \approx 2000$, $\snr \sim 4$). 
In the near future, the Baryon Oscillation Spectroscopic Survey (BOSS,
part of SDSS-III\footnote{\url{http://www.sdss3.org}})
aims to increase the sample size of high-redshift ($z \gtrsim 2$) quasars to 
$\gtrsim 10^5$ at a similar spectral resolution to the SDSS.

Many of the techniques (Voight profile fitting, wavelet analysis etc.) 
developed for probing the IGM are unsuitable for use with the lower quality of the
SDSS data, while the flux statistics that have been
 measured for the SDSS \lya\ forest, such as the flux power spectrum, are relatively
insensitive to the detailed astrophysics of the IGM. 
For example, \citet{lai+06} have shown that the SDSS flux power spectrum is insensitive to
large-scale temperature inhomogeneities arising from \heii\ reionization.

In this paper, we borrow some statistics used to measure 
 mechanical and transport properties of disordered media in the material sciences,
which we term the `threshold probability functions',
and explore their applications to mock \lya\ forest spectra based on the SDSS sample.
We will generate simple toy models for the IGM based on different scenarios for 
\heii\ reionization, and explore the ability of the
threshold probability functions to distinguish between them. 

We first introduce and define these statistics in \S\ref{sec:defin}, before digressing
to discuss the simulations and toy models we use to generate the mock spectra in 
\S\ref{sec:model}. In \S\ref{sec:pdf} we calculate the flux PDF as a check on our errors, 
before going on to apply the threshold statistics on the mock data in \S\ref{sec:stwo}.

\section{Definition of Threshold Statistics:\\ \stwo, \ctwo, and \dtwo}
\label{sec:defin}

In the study of the \lya\ forest, the two-point flux correlation 
function is commonly defined as:
\begin{equation}\label{eq:fluxcf}
\xi(r) = \langle \delta_F(r')\: \delta_F^{*}(r' + r) \rangle,
\end{equation}
where $r$ is the comoving distance  
between two points in the line-of-sight of the background quasar
(equivalently expressed as redshift, wavelength, or velocity intervals within
the observed spectrum), and
\begin{equation}
\delta_F(r) = \frac{F(r)}{\bar{F}} - 1 
\end{equation}
where $F(r) = e^{-\tau(r)}$ is the flux transmitted through the \lya\ optical depth $\tau(r)$ 
at a given point, and $\bar{F}$ is the 
mean transmitted flux in the spectrum. 
Another commonly used statistic is the Fourier transform of 
$\xi_F(r)$, the flux power spectrum
\begin{equation}
P_F(k) = \int^{\Delta r/2}_{-\Delta r/2} \xi_F(r') e^{ikr'} dr' 
\end{equation}
computed over the interval $\Delta r$, and $k$ is the wavenumber.

In this paper we introduce to astrophysics several related two-point
statistics used in material science \citep{ torq+88,jiao_torq09}. 
For a volume that is occupied by a two-phase medium, 
for any one of the phases (say phase $i$) we can define
\begin{equation}\label{eq:stwodef}
\stwohat(\vec{r_1}, \vec{r_2}) = \ctwohat(\vec{r_1}, \vec{r_2})
+\dtwohat(\vec{r_1}, \vec{r_2}).
\end{equation}
\stwohat\ is the probability function of finding both points $\vec{r_1}$
 and $\vec{r_2}$ in phase $i$, while \ctwohat\ and \dtwohat\ denote the probability 
of finding points of the phase $i$ within the same `cluster'
(in this paper referring to contiguous groupings of pixels, {\bf not} galaxy or star clusters),
and in separate clusters, respectively.
This is illustrated in Figure~\ref{fig:stwodef}. 
\ctwohat, known as the `cluster function', encodes higher-order
not contained in two-point statistics \citep{jiao_torq09}. 
For astrophysical purposes we assume an isotropic and homogeneous field,
so $\stwohat(\vec{r_1}, \vec{r_2}) = \stwohat(|\vec{r_1}- \vec{r_2}|) = \stwohat(r)$ etc.

% Figure illustrating S2 etc in a spectrum. 

We generalize these statistics for use with the \lya\ forest by defining them as a function
of flux threshold, \fth. At each value of \fth, we divide the spectrum into high
($F > \fth$) and low ($F < \fth$) regions and then compute the clustering properties 
for these phases. In this paper we focus on the high ($F > \fth$) phases which are more sensitive
to \heii\ reionization. This defines two dimensional functions, $\stwohat(r, \fth)$,
$\ctwohat(r, \fth)$, and $\dtwohat(r, \fth)$.

\bfig
	\epsscale{1.15}	
	\plotone{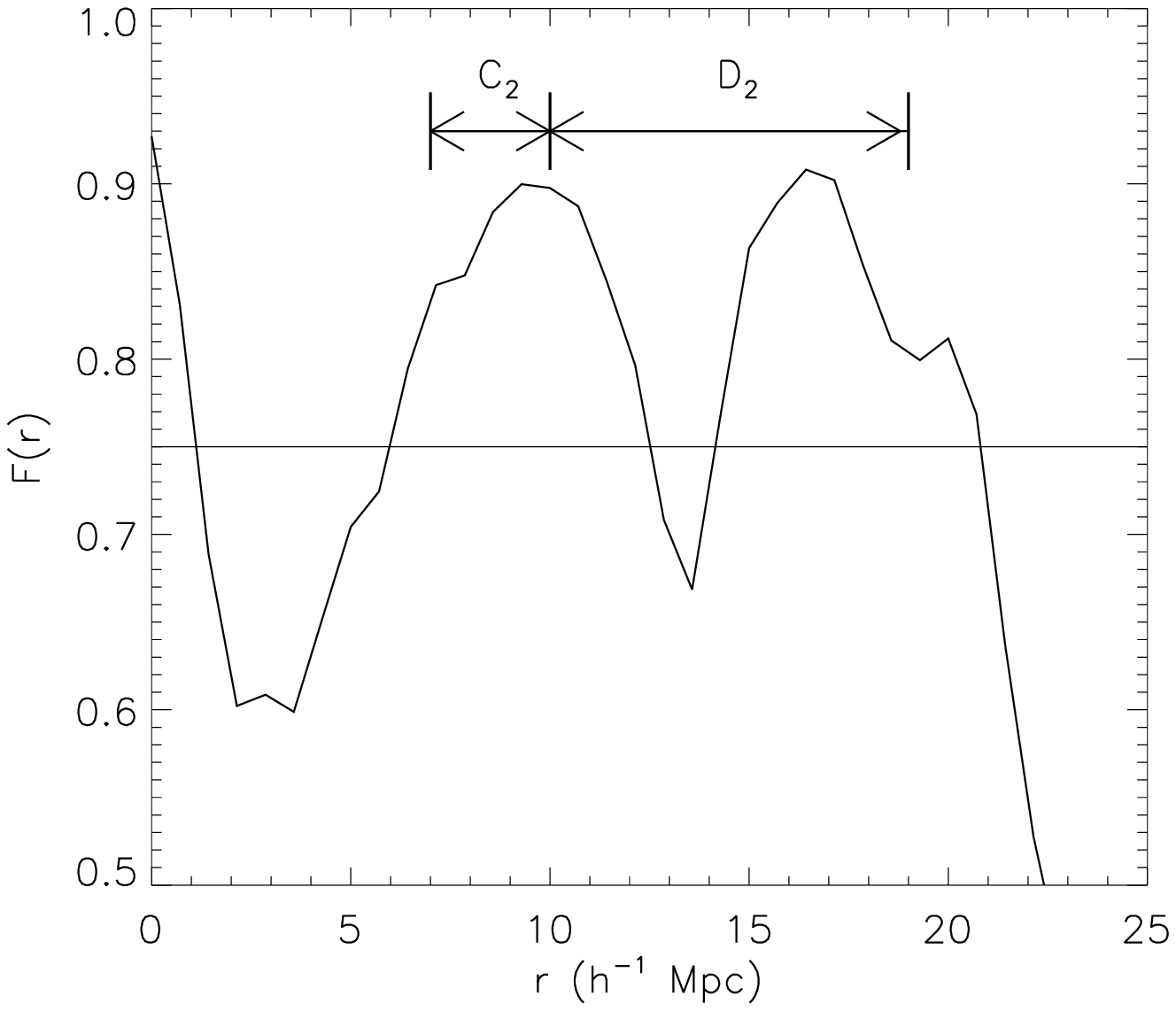}
	\caption{\label{fig:stwodef}A segment of transmitted \lya\ forest flux plotted as a function of
	comoving distance $r$ along the line-of-sight,  illustrating
 	the definition of the threshold probability functions
	\ctwohat\ and \dtwohat.
	All pixels above the flux threshold ($\fth = 0.75$ in this example)
	 contribute to \ctwohat, \dtwohat, and $\stwohat \equiv \ctwohat + \dtwohat$. 
	The arrows at left illustrate a pixel pair within the same `cluster' of pixels which
	contribute to \ctwohat, while the arrows at right denote another pixel pair in different clusters
	which contribute to \dtwohat. 
		}
\efig

The threshold probability function at zero lag, $\stwohat(0 |F_{th})$ is 
directly related to the familiar flux probability distribution function (PDF), $p(F)$:
\be \label{eq:s20}
\stwohat(0 |F_{th}) = \int^1_{F_{th}} p(F) dF.
\ee

The number of possible pixel pairs 
(and hence \stwohat) decreases linearly with $r$ within a finite length $L$.
In order to take this effect into account we introduce a rescaled version of \stwohat:
\be
{\stwo}(r|\fth) = \stwohat(r|\fth) \left( 1 - \frac{r}{L} \right), 
\ee
and similarly for \ctwohat\ and \dtwohat. 
At large separations, $S_2(r | F_{th} ) \rightarrow S^2_2(0 |F_{th})$ if there is
no large-scale order in the system. 
We refer to $\stwo(r | \fth)$, $\ctwo(r | \fth)$, and $\dtwo(r | \fth)$ collectively as the
`threshold probability functions'. 
Intuitively, they can be thought of as the flux
PDF evaluated as a function of correlation length.

These threshold probability functions should not be confused with the threshold 
crossing statistics that counts the number of times that the observed \lya\ forest
transmission spectrum intersects a given flux value per unit redshift 
\citep{mirald+96, fan+02}. The threshold crossing statistic is a generalization
of the forest line density and does not contain spatial information, unlike
the threshold probability functions. 

% More identities

While \heii\ reionization may create a
 two-phase thermal structure in the IGM
 \citep{lai+06,mcquinn+09}, its imprint
 on the \lya\ forest will {\bf not} be manifested as distinct phases
in the flux distribution of the \lya\ forest spectra.
It is the density field which predominantly determines the distribution of low-
and high-absorption regions. These thermal inhomogeneities, however, 
do modulate the amplitude of the optical depth and leave a potentially
statistically detectable effect especially in the low-absorption regions.
It is this effect which we are attempting to detect using the threshold 
clustering functions.

\section{\lya\ Forest Model}\label{sec:model}

In this section, we describe a series of simulations of the \lya\ forest 
that we will use to test
ability of the threshold probability functions to distinguish between 
different \heii\ reionization histories.

\subsection{Simulations}

We use a set of publicly available\footnote{\url{http://mwhite.berkeley.edu/BOSS/LyA/Franklin/}} 
 mock \lya\ forest spectra \citep{slosar+09} that have been
generated from dark matter-only particle-mesh
simulations based on a flat $\Lambda$CDM cosmology with 
$\Omega_{\Lambda}=0.75,\Omega_M = 0.25, h=0.75, n=0.97,$ and $\sigma_8=0.8$.
The simulations evolved $3000^3$ particles in a $1500 \mpc$ box with
the forces computed on a $3000^3$ grid. Density and velocity fields were 
then generated using spline-kernel interpolation with an effective smoothing 
radius of $250 \kpc$, and line-of-sight skewers extracted at redshift $z=2.5$ 
with a spacing of $10\mpc$ to provide $150^2=22500$ skewers from each 
run.

The \lya\ optical depth in each pixel was then generated using the 
fluctuating Gunn-Peterson approximation \citep[FGPA, ][]{croft+98, gned+hui97}:
\be \label{eq:fgpa}
\tau \propto \bar{T}^{-0.7} \Delta^{2-0.7(\gamma-1)} 
\left(1 + \frac{1}{H(z)}\frac{dv_{pec}}{dx}\right),
\ee
where $\Delta=\rho/\bar{\rho}$ is the density perturbation, $H(z)$ is the
Hubble parameter at redshift $z$, ${dv_{pec}/dx}$ is the peculiar velocity
gradient, $\bar{T}$ is the temperature at mean density, and $\gamma$ 
parametrizes the equation of state between density and temperature (see Equation~\ref{eq:eos}).
In these runs $\bar{T}= \scien{2}{4} K$ and $\gamma = 1.5$ was assumed. 
 In addition to the
optical depth skewers, matching skewers of the underlying density were also made available.  

These dark matter simulations are not expected to accurately capture the small-scale
power of the \lya\ forest as hydrodynamics are not included. 
However, they provide a fiducial set of mock spectra from which it is
easy to generate different toy models of the IGM by adjusting the optical 
depths using the FGPA (Eq.~\ref{eq:fgpa}).
This provides a convenient testing ground for the ability of the threshold probability functions
to distinguish between differences arising from different IGM models.

\subsection{Homogeneous IGM Models}
% Talk about why \gamma =1.5 is expected and describe obs that say \gamma<0
The fiducial set of simulated spectra described above
assumes a homogeneous\footnote{Note that in this paper `homogeneity' 
and `inhomogeneity' refer to the spatial distribution
of the IGM  thermal properties, primarily $\gamma$ and/or $\bar{T}$.} IGM in which 
$\bar{T}=\scien{2}{4} K$ and $\gamma=1.5$.
This is the value of $\gamma$ at which unshocked gas settles
at $z \sim 3$ after a hydrogen reionization event at $z \gtrsim 6$  \citep{hui+gned97},
while $\bar{T}=\scien{2}{4} K$ is approximately the value obtained through
line-profile fitting of high-resolution \lya\ forest spectra \citep{mcd+01,theuns+02}. 
We refer to this fiducial model as `G1.5'.

We study the effects of changing the equation of state uniformly across the IGM by creating
models with $\gamma$ set to 1.3 and 0.8 (models G1.3 and G0.8, respectively). 
$\gamma = 0.8$ represents an inverted equation of state predicted by \citet{furl+oh08-2}
for scenarios in which dense regions, which were reionized early on, have had time to cool adiabatically
to temperatures lower than more recently reionized voids.
We introduce G1.3 as an intermediate case between G1.5 and G0.8, although as we shall 
see later, it has a flux PDF which is degenerate with our inhomogeneous reionization models.

The mean temperature $\bar{T}$ is kept unchanged, 
as global variations in temperature will be manifested as changes in 
the mean flux level\footnote{This is not entirely true as thermal broadening
would change the small-scale power of the \lya\ forest, but due to the 
 low-resolution of our simulations and the fact that we are 
looking for large-scale variations, we ignore this effect.}, 
which we regard as a fixed parameter by normalizing
our \lya\ forest spectra to the same value of $\langle F \rangle = 0.8$ \citep{meik+white04}.

\subsection{Inhomogeneous IGM Models}

\begin{figure*}
\bc
\includegraphics[width=0.33\textwidth]{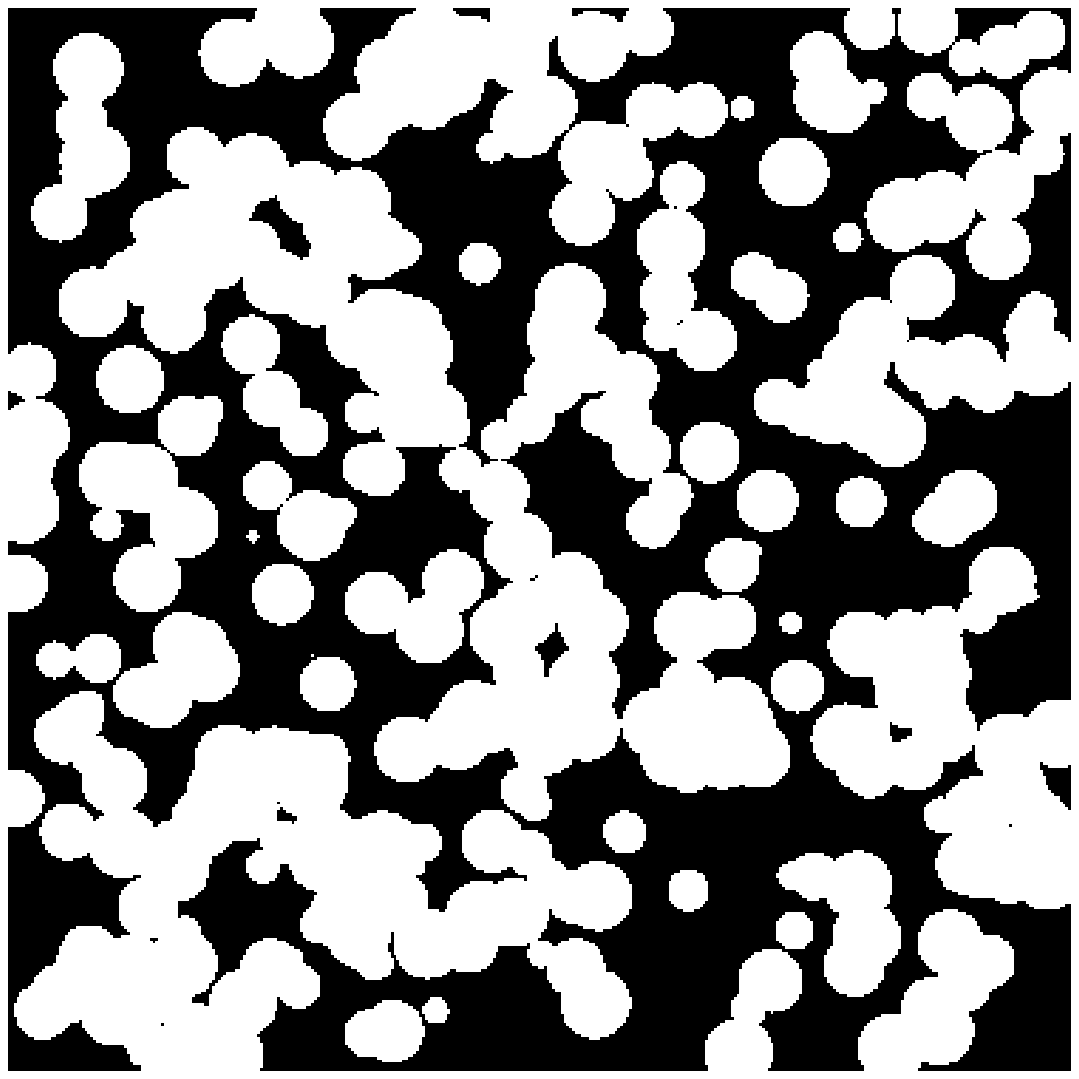}
\includegraphics[width=0.33\textwidth]{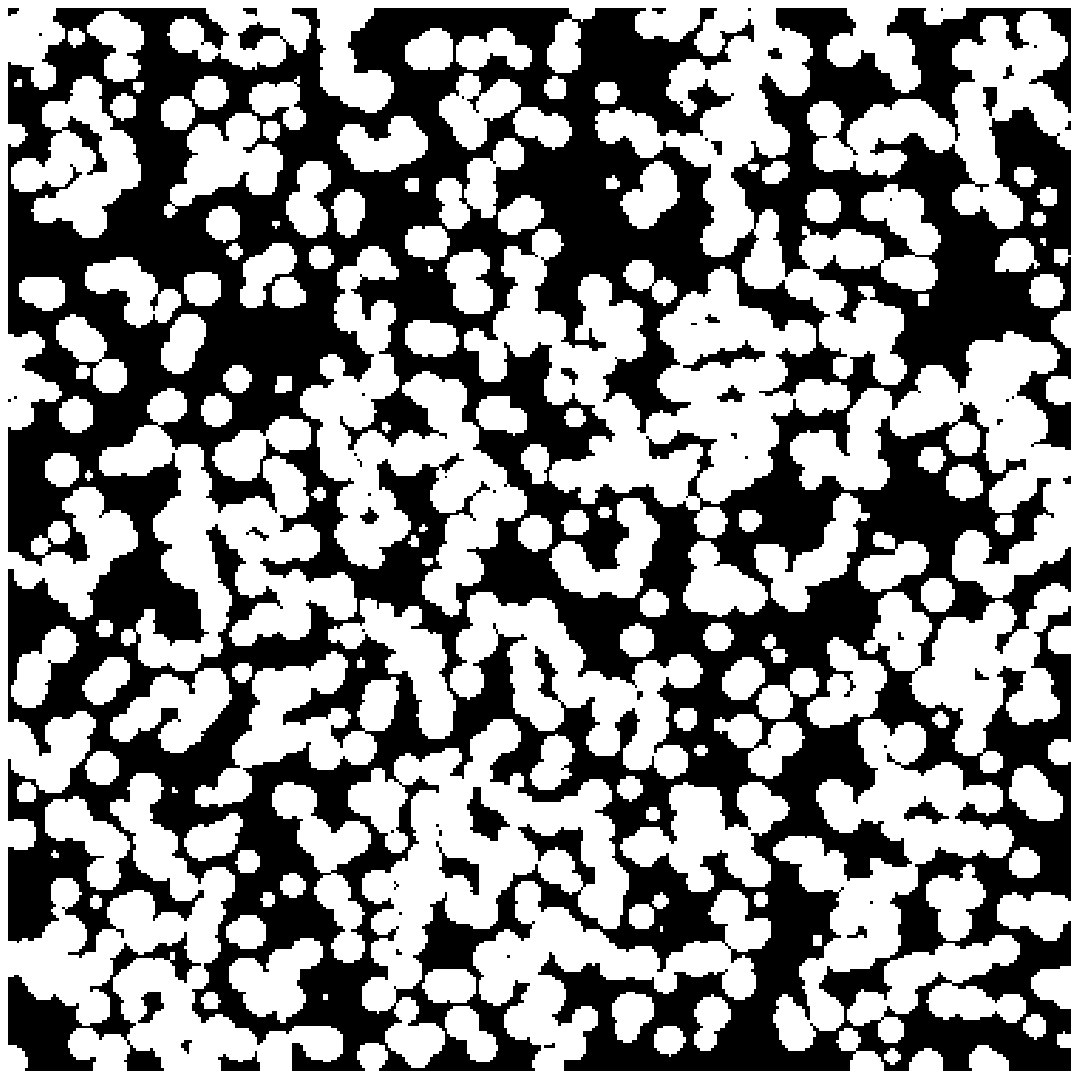}
\includegraphics[width=0.33\textwidth]{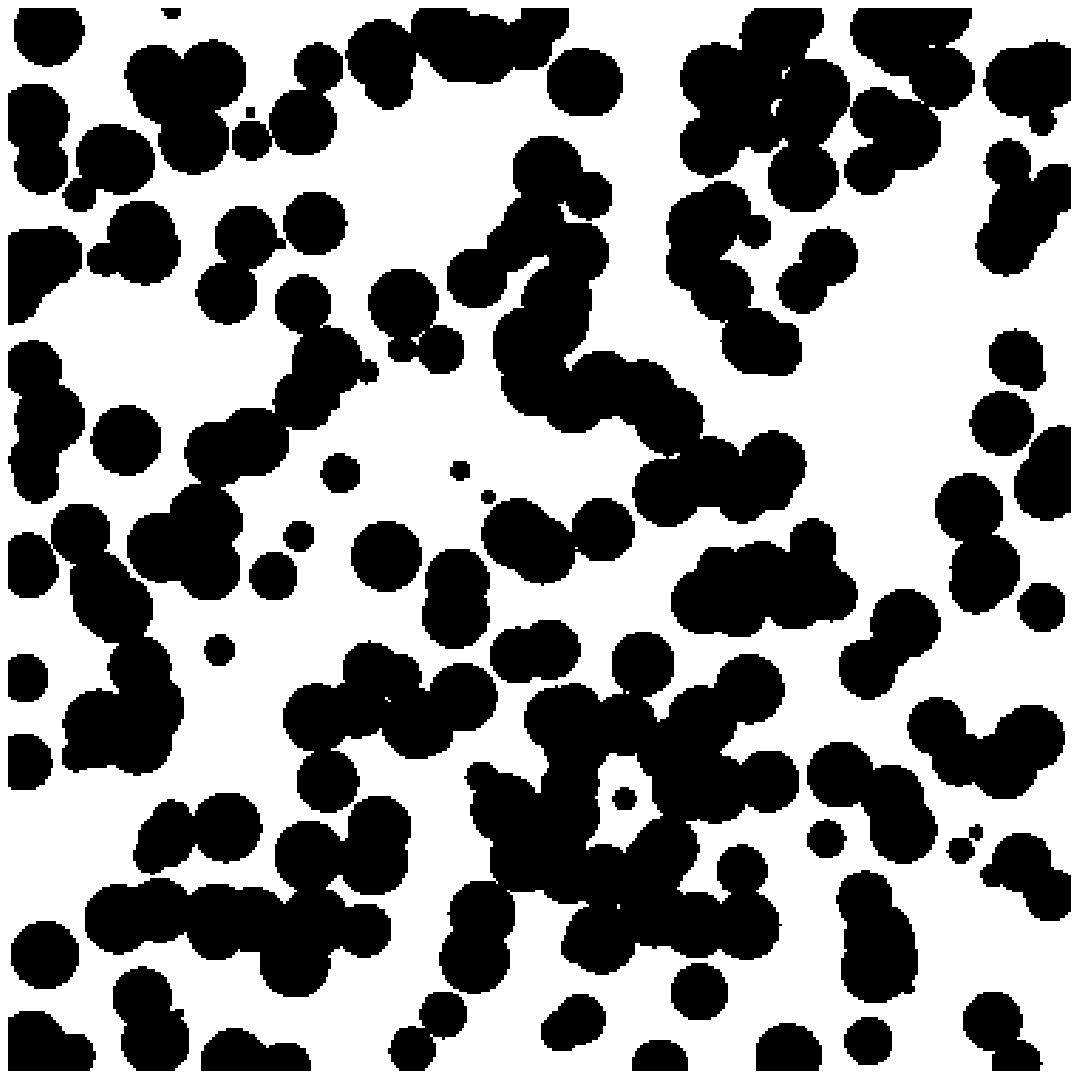}
\ec
\caption{Examples of
pixel masks used to generate 
the inhomogeneous IGM models, shown here as 2-dimensional cross-sections
across the box which is 1500 \mpc\ on a side. From left to right: Models R50, R25, and I50. 
White represents the hot IGM phase
while black is the cold phase with both occupying 50\% of the volume. 
The mock spectra will thus represent 1-dimensional
samplings through the hot and cold phases shown here.
 \label{fig:bubbles}}
\end{figure*}

Several authors \citep{lai+06,mcquinn+09} 
have suggested that the reionization of \heii\ by quasars at $z \sim 3$
results an inhomogeneous IGM with significantly heated regions. 
In this paper we consider a basic picture of inhomogeneous \heii\ reionization 
inspired by the simulation results of \citet{mcquinn+09}, in which the IGM is composed of 
hot and cold phases with equal volume-filling factors of 50\% each. The hot phase
has a mean temperature $\bar{T}_h = \scien{2.5}{4} K$ and $\gamma=1.2$, 
while the cold phase has a mean temperature $\bar{T}_c = \scien{1.5}{4} K$ and $\gamma=1.5$.
Note that while changes in the overall mean temperature will be absorbed
in the mean flux level, the contrast between $\bar{T}_h$ and $\bar{T}_c$ is
more important as it
results in relative changes of optical depth independent of the overall flux normalization.
% Mention differences between this toy model and others

We approximate the spatial distribution of the hot and cold regions by 
generating a pixel mask in which bubbles of radius $R_{bub}$ are randomly inserted
into a 3-dimensional box with a comoving volume $(1500 \mpc)^3$,
the same size as the simulation box.
These bubbles, which are allowed to overlap, are added until
 50\% of the volume lies within bubbles. 
This allows a full distribution of intersection path lengths within the mock spectra, 
from close to zero to several times $R_{bub}$ as illustrated in Figure~\ref{fig:bubbles}.
Our simple model provides a simulacrum of the complex 
spatial morphologies seen in \citet{mcquinn+09}. 
Note that we neglect peculiar velocities 
as the bubble positions are completely random and have
 no overall correlations with large scale-structure.

The mock spectra from the simulations are then modified on a pixel-by-pixel basis 
depending on whether the three-dimensional spatial position of each pixel was flagged by 
the pixel mask: the optical depths of pixels which fall within hot bubbles are rescaled using
the FGPA (Equation~\ref{eq:fgpa}) so that they have $\bar{T} = \bar{T}_h$ and $\gamma=1.2$, 
while pixels outside the bubbles have $\bar{T} = \bar{T}_c$ and $\gamma=1.5$.
Note that this does not take into account the effect of thermal broadening. 
Although thermal broadening primarily affects the small-scale power of the 
forest, it also has a small effect on the large-scale bias. 
This coupling between the large- and small-scale power is not well-understood 
and will need to be accounted for
in a full data analysis using large-scale hydrodynamical simulations, but
for the approximate treatment in this paper we ignore this effect.

The characteristic scale of the thermal inhomogeneities arising from \heii\ reionization
are expected to be dependent on quasar physics such as duty cycles, clustering and ionizing spectrum
\citep{mcquinn+09}. We thus test the threshold probability functions to this characteristic scale by 
introducing models with hot bubbles of $R_{bub}=50 \mpc$ and $R_{bub}=25 \mpc$, denoted 
by R50 and R25 respectively. Both these models have the same volume filling fraction of 50\% for 
hot bubbles.

In addition, we created a model, I50, in which the topology of \heii\ reionization is inverted, 
i.e.\ the IGM is composed of {\it cold} bubbles of $R_{bub}=50 \mpc$ with the surrounding
regions filled with the hot phase, with the properties of the hot and cold phases identical
to those of models R50 and R25. Although this model is not physically realistic, we include it
to test the sensitivity of the threshold statistics towards topology. 
Figure~\ref{fig:bubbles} illustrates the inhomogeneities in our various models.

All the inhomogeneous models, R50, R25, and I50, have the same total number
of pixels intersecting with hot regions when averaged across large numbers of spectra. 
The differences lie in the manner the pixels are spatially distributed 
along the skewers.

\subsection{Mock Spectra}
 
\begin{figure*}
%\epsscale{0.85}
\plotone{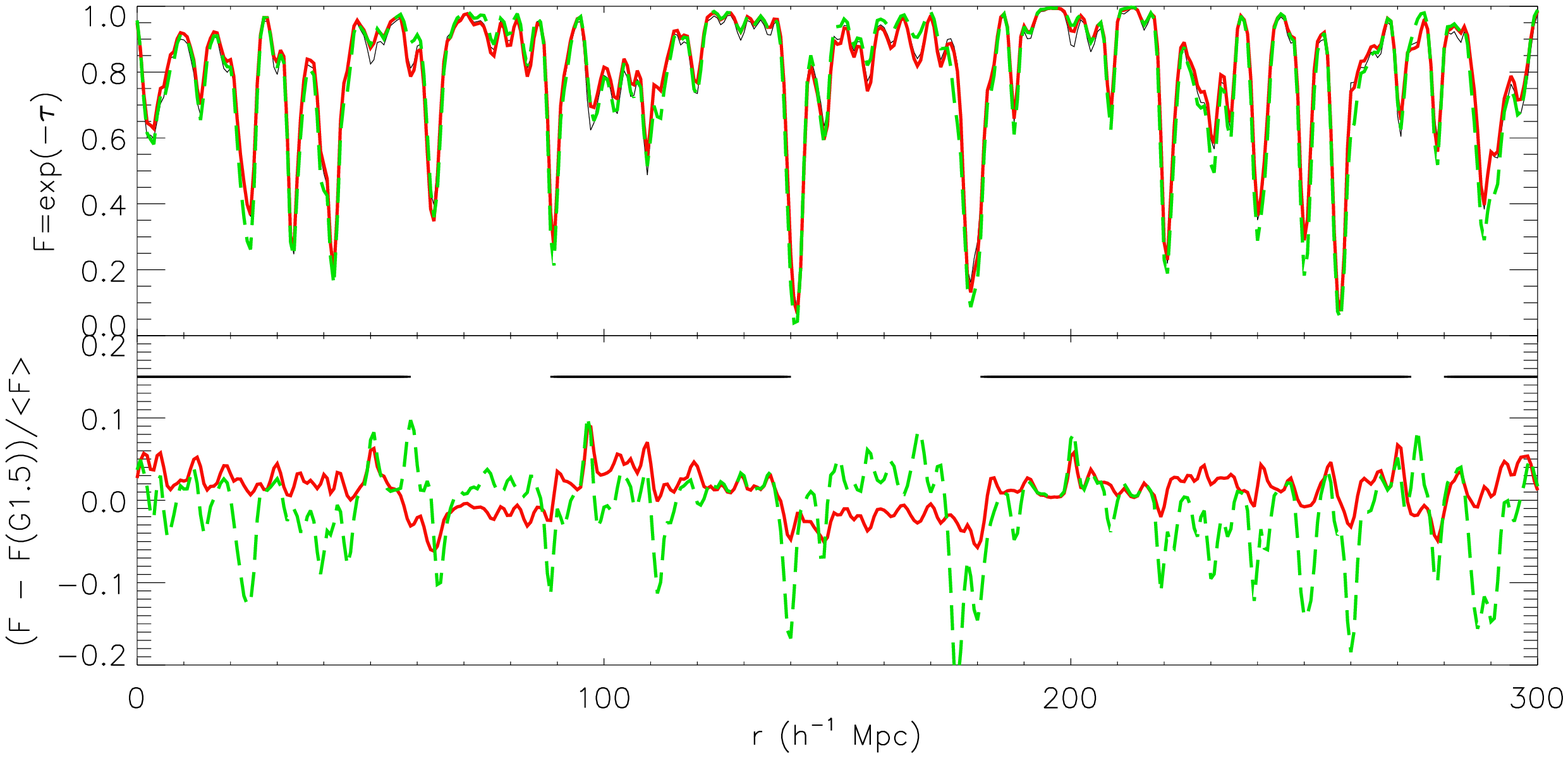} 
\caption{\label{fig:fluxplot} Top panel: Flux transmission spectra from models G1.5 (thin black line), 
G0.8 (green dashed line), and R50 (thick red line) for one line-of-sight
-- shown here without noise or other
systematics. Bottom panel: 
Difference in flux between models G0.8 and G1.5 (green dashed line), and
between R50 and G1.5 (solid red line), normalized by the mean flux. 
The horizontal bars indicate the portions of the line-of-sight which 
intersect with hot bubbles. See electronic edition of the Journal for a color version of this figure.}
\end{figure*}

As the primary purpose of this paper is to study the ability of the 
threshold statistics to constrain the IGM using existing SDSS data, 
in this paper we model the instrumental and systematic
effects at an approximate level sufficient to estimate the errors.
A more detailed approach is to be deferred to a later
data analysis paper. 

First, flux transmission spectra are generated from the
optical depth skewers from $F = \exp{(-\tau)}$, with a
global mean flux set to $\langle F \rangle = 0.8$.
 The spectra are then Gaussian-smoothed to the resolution 
of the SDSS spectra which have FWHM=$150 \kms $ 
 (corresponding to $\simeq 1.4 \mpc$ at $z=2.5$)
 and binned into $70 \kms$ pixels.
 We then split the $1500 \mpc$ simulation skewers 
 into segments of length $500 \mpc$, 
approximately the comoving distance subtended by individual \lya\ forest
lines-of-sight at $z=2.5$. 

In the upper panel of Figure~\ref{fig:fluxplot} we show the 
 flux transmission for a segment of noiseless
mock \lya\ forest spectrum in the fiducial simulations (model G1.5), along with 
corresponding spectra for models G0.8 and R50.  
It is difficult to distinguish between the different cases by eye 
in Figure~\ref{fig:fluxplot}, although G0.8
 shows more contrast between the high- and low-absorption regions, which is due
 to the stronger scaling of optical depth with density 
 ($\tau \propto \Delta ^{2-0.7(\gamma-1)}$) when $\gamma$ is decreased.

The lower panel shows the difference in flux relative to model G1.5, 
divided by the mean flux. For model G0.8 the increased absorption occurs
within the \lya\ lines where the optical depth is high. 
The flux in R50 shows a slight overall increase ($\sim 5\%$)
within regions which intersect hot bubbles, relative to the cold regions. 
While the increased transmission from the bubbles can be picked out by eye from 
the difference spectra in Figure~\ref{fig:fluxplot}, 
{\it a priori} it would be impossible to identify these regions from
a given spectrum.

As the SDSS spectra are relatively noisy, the mock samples need
to have approximately the correct noise and sample properties.
We estimated the signal-to-noise ratio per pixel ($\snr$)  in the
\lya\ forest of the SDSS data in following manner: from the latest
SDSS Data Release 7 (DR7) quasar catalog \citep{schneider+10} we selected quasars in the
redshift range $2.4 \leq z_{qso} \leq 2.7$.
In the wavelength region $(1 + z_{qso})1040 \AA \leq \lambda \leq (1 + z_{qso})1180 \AA$
(the typical range of usable \lya\ forest), the $\snr$ for each sightline 
is then estimated as the median of $f_{\lambda,i}/\sigma_{N,i}$ (where $f_{\lambda,i}$
 and $\sigma_{N,i}$ are the observed flux and pipeline noise
from the individual pixels).
While the absorption of metals such as \ion{C}{4} and 
\ion{O}{6} are not included in this simplified model, a full analysis of actual SDSS spectra
should use metal absorption measurements from high-resolution data to estimate 
this systematic.

Within the aforementioned redshift range there are
3217 quasars in DR7, of which 1641 have $\snr \geq 4$ in the \lya\ forest.
We thus set our mock sample size to be 1500 spectra with $\snr=4$ per pixel, 
in which Gaussian noise with variance $\sigma^2_N = (\overline{F}/\snr)^2$
was added to each pixel in the mock spectra to simulate the 
instrumental noise, 
where $\overline{F}$ is the mean flux within each skewer. 

While the instrumental modelling is carried out at a relatively simple level in this paper, 
the actual instrumental systematics of the SDSS data are well understood 
\citep[e.g.][]{stoughton+02, mcd+06} and should be included in any analysis of real data.
 The biggest known unknown 
 is the ability to accurately to fit the quasar continuum.

\section{Quasar Continuum Fitting And Flux PDF}
\label{sec:pdf}

Since the threshold probability functions are evaluated as a function of 
the transmitted flux in the \lya\ forest, 
they are sensitive to the fitting of the intrinsic quasar continuum.
The fitted continuum fixes the zero-absorption level and determines the
normalization of the flux transmission in the \lya\ forest. 
The uncertainties in continuum fitting are likely the dominant
source of systematic error and are not as well understood as the 
SDSS instrumental noise.

 In this section we discuss 
the continuum-fitting methods that have been published in the literature
as well as their associated errors, and compute the flux PDF as a check on our assumed
errors. We will see that the flux PDF from the SDSS data in itself can put interesting
constraints on the IGM.

\subsection{Continuum Fitting}
\citet{desj+07} carried out a study of the systematics involved in measuring the 
 \lya\ forest flux PDF from the
SDSS (albeit from the earlier Data Release 3 (DR3)), by comparing it with mock
spectra generated using parameters derived from high-resolution \lya\ forest spectra. 
They found that the discrepancies between their mock spectra and the
measured SDSS PDF can largely be accounted for by errors in the quasar continuum fits. 
Individual forest spectra were with a power law and Gaussian curves for the 
quasar emission lines (e.g.\ at $1070 \AA$ and $1120 \AA$ in the quasar restframe), 
a technique first introduced by \citet{bern+03} for use on composite quasar spectra.

\citet{desj+07} estimated an error of $\sigma_F \approx 20\%$
 in their determination of the individual quasar continua, although this was derived from 
 the error budget of the measured flux PDF rather than from a detailed analysis
  of individual fitted continua.

\citet{suz+05} discussed quasar continuum fitting using principal component
analysis (PCA) based on $\sim 50$ ultraviolet quasar spectra obtained from the 
Hubble Space Telescope.
At the low-redshifts ($z\sim0.5$) of their data set there is little \lya\ forest absorption,
which allows the quasar continuum to be accurately measured in wavelength regions
which normally suffer from considerable \lya\ forest absorption.
They reported a typical error of $\sigma_F \approx 9\%$
in estimating the \lya\ forest continuum 
 using only points red-wards of the \lya\ emission line 
which would be unaffected by the \lya\ forest. 
In addition, they found that the PCA gave good fits for the shape of the quasar
continuum at the \lya\ forest wavelengths, 
even when the amplitude was not accurately predicted from the red-side of the 
spectrum. 

While the PCA fitting method on low \snr\ spectra needs to be extensively tested,
we expect it to be more accurate than the power-law fits as it would in principle account
much of the large-scale structure in the intrinsic quasar continuum
arising from weak emission lines, which could 
be degenerate with large-scale inhomogeneities in the IGM.
For the purposes of our mock spectra we take the findings of \citet{suz+05} at face-value,
and adopt a model for the continuum fitting errors in which the shape of the 
continuum is assumed to be perfectly predicted, leaving only a constant normalization error
with no tilt or wiggles in the residual. 
The errors in the continuum level are then obtained by normalizing each 
individual mock spectrum by a local mean flux $\fbar$ 
drawn from  a Gaussian distribution with a global mean flux $\langle F \rangle = 0.8$ 
\citep{meik+white04}, and standard deviation $\sigma_F = 9\%$. 

\subsection{Flux PDF}\label{subsec:pdf}
\begin{figure*}
%\epsscale{0.85}
% Generated from 'pdfsig' and 'pdfdiff' in tau_toy/pdf_cosvar/plotnoise.pro
\plottwo{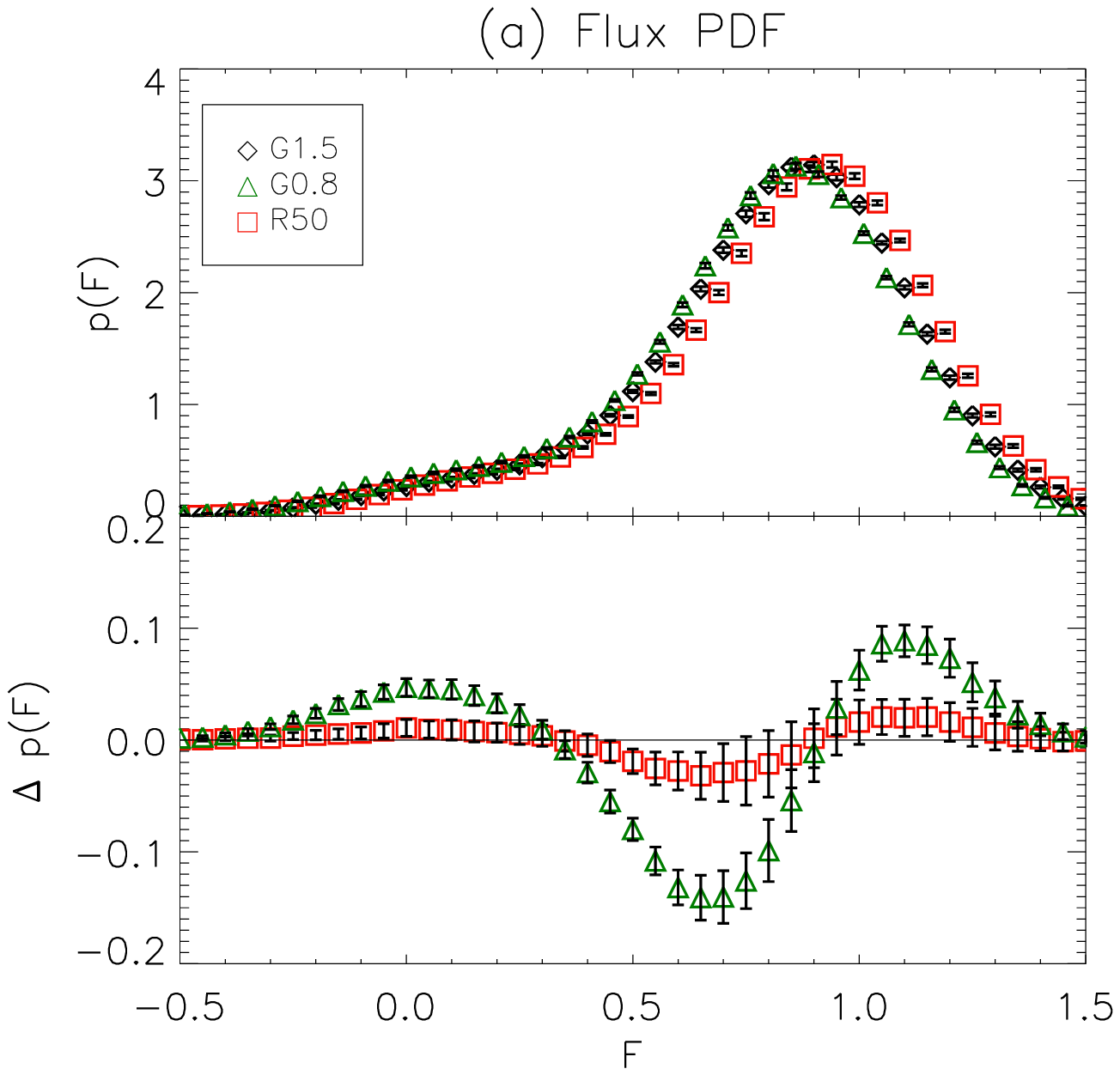}{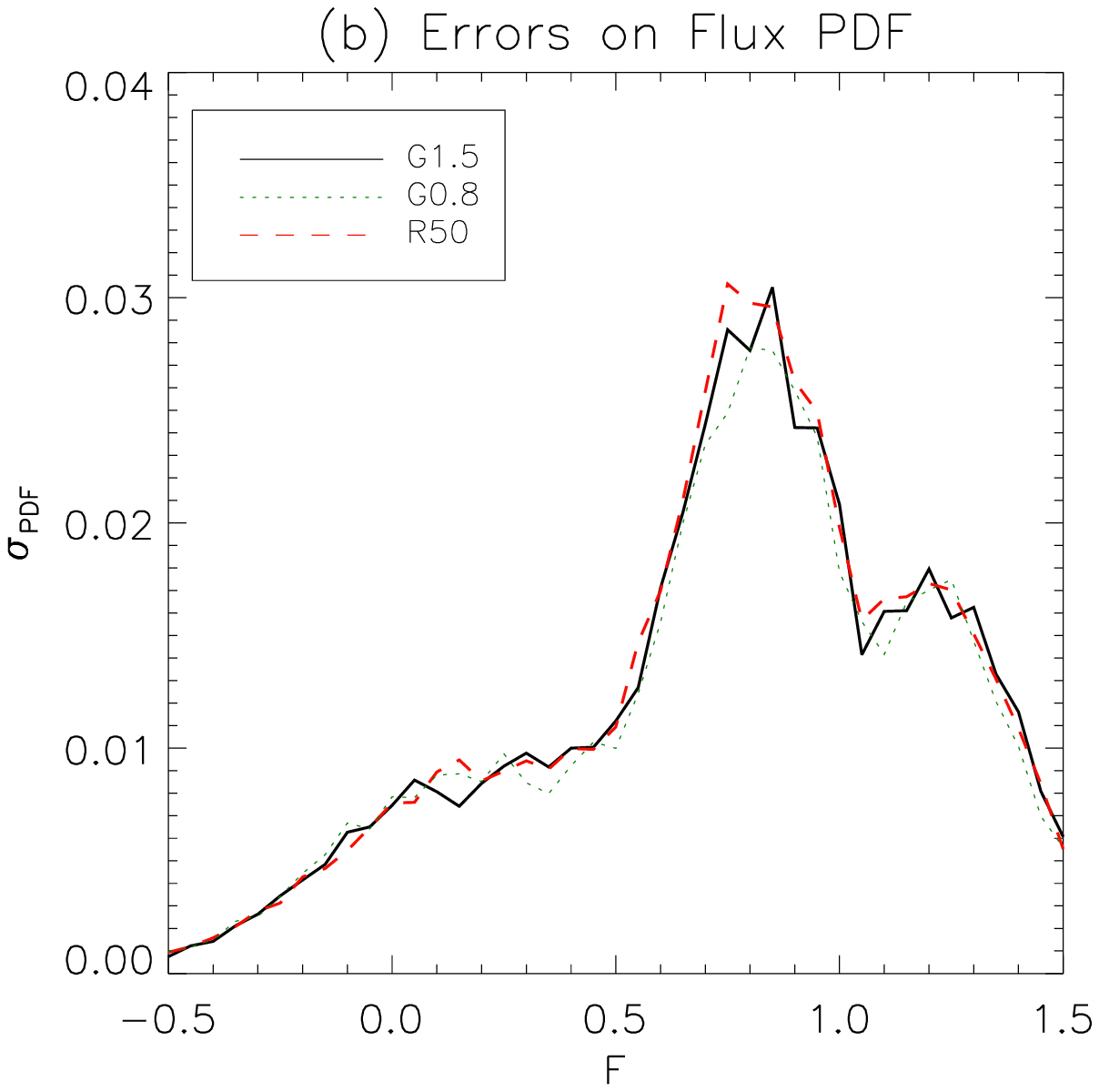}
\figcaption{\label{fig:pdf} (a) Top panel shows the
 flux PDF from 600 mock spectra with $\snr=4 $ per pixel at $z = 2.5$, generated for
models G1.5 (black diamonds), G0.8 (green triangles), and R50 (red squares). 
The points are offset horizontally for clarity. Lower panel shows the difference
in the PDF between models G0.8 and G1.5 (green triangles), and between 
models R50 and G1.5 (red squares).  
 (b) Estimated errors on the flux PDF
 from models G1.5 (black solid line), G0.8 (green dotted line), and
R50 (red dashed line). See electronic edition of the Journal for a color version of this figure.}
\end{figure*}

In order to provide a comparison for our assumed systematics 
vis-\'{a}-vis \citet{desj+07}, 
we evaluate the PDF for our toy models
computed from mock samples similar to the DR3 data, 
i.e.\ 600 spectra with $\snr=4$ per pixel at $z\sim 2.5$.

The resultant PDFs for models G1.5, G0.8, and R50, 
are shown in the top panel of Figure~\ref{fig:pdf}a, with each PDF 
divided into bins of width $\Delta F = 0.05$.
The error bars are the $1\; \sigma$ dispersion from $\sim 100$ realizations.
These realizations have different random seeds for the instrumental noise, 
and cosmic variance
is included by drawing each realization from different lines-of-sight in the
simulation box and generating the bubble distributions separately in the case
of the inhomogeneous models.
The errors are plotted explicitly in Figure~\ref{fig:pdf}b.

The PDFs from our mock spectra have the general characteristics expected
from noisy \lya\ forest spectra, e.g.\ the existence of pixels with 
$F < 0$ and $F > 1$. 
The fact that the plots in Figure~\ref{fig:pdf}a are clearly different from the 
PDFs of high-resolution spectra e.g.\ in \citet{kim+07}
 underlines the fact that the flux PDF is sensitive to noise and other 
systematics. 
We do not expect our PDF to match the observed SDSS PDFs in \citet{desj+07} 
exactly as our mock
spectra are generated from simulations that do not have the right small-scale power, 
nor did we carry out a detailed consideration of the systematics such as instrumental noise.
Such a careful approach would be required when analyzing real data in order to 
constrain {\it a priori} the overall \lya\ forest parameters, but in this paper we are concerned
with the differences arising from the IGM with respect to a fiducial model, so the fiducial model
itself does not need to be exactly right for our tests.

We use the errors on the PDFs as a check on our assumptions
regarding the systematics. 
As expected, increasing the scatter $\sigma_F$ of the continuum levels
increases the error bars on the PDF.
We find that the errors from our mock PDFs, Figure~\ref{fig:pdf}b,
 and those of 
\citet{desj+07} (Figure~7 in their paper) are similar when we use  
$\sigma_F = 9\%$ as suggested by \citet{suz+05}.
This indicates that our simplified models for the SDSS instrumental effects
 and sample properties 
provide a reasonable estimate for the sample errors.
For our DR7 mock samples in the later parts of this paper 
we will henceforth assume that the shape of the quasar
continuum can be fit perfectly, and set errors in the continuum level
to $\sigma_F = 9\%$.  

The lower panel of Figure~\ref{fig:pdf}a shows the differences in PDF with respect to the fiducial
model G1.5, $\Delta p(F)_\mathrm{model} = p(F)_\mathrm{model} - p(F)_{G1.5}$. 
Interestingly, it is clear that even with the DR3 data set it would
have been possible to distinguish an inverted equation of state, G0.8, 
from the fiducial model G1.5 at high significance.
 
In order to quantify the ability of the PDF to distinguish between the different models, we compute a logarithmic likelihood
\be \label{eq:like}
\lnl(\mu_i|x_i) = \frac{1}{2} (x_i - \mu_i)^{T}C_{ij}^{-1}(x_j - \mu_j),
\ee
 where we assume the PDF for one model, $x_i$ 
-- the subscript refers to the bins
in the data --,  to be the `observed' data 
and take the mean PDF for another model $\mu_i$ as the `theory' points, and 
$C_{ij}$ is 
the covariance matrix for $x_i$ (in this case directly evaluated
from the mock realizations).
A small value of $\lnl \sim 1$ indicates that the `theory' model is 
consistent with the `observed' model and hence the two models 
cannot be differentiated.
 As the estimated errors are similar for all
the models as shown in Figure~\ref{fig:pdf}b, in practice $x_i$ and $\mu_i$ 
are interchangeable for any two models. 
When evaluating Equation~\ref{eq:like} on the flux PDF 
we use only bins in the range $0.4 \leq F \leq 1.1$, which 
excludes the lower and upper 10\% of the flux distribution.

Table~\ref{tab:pdflike} summarizes the results from assuming a DR7 sample size
(1500 quasars) and the systematics we have assumed. Each entry in the table shows \lnl
for distinguishing between PDF of the `observed' model in the corresponding column,
 and the PDF of `theory' model in the corresponding row.

\begin{deluxetable}{l|cccccc}
\tablewidth{0pt}
\tablecaption{\label{tab:pdflike} \lnl between PDFs for different IGM models}
\tablehead{ & \colhead{G1.5} & \colhead{G1.3} &\colhead{G0.8} &\colhead{R50} &\colhead{R25} &\colhead{I50} }
% Computed in 'pdflike' in tau_toy/pdf_cosvar/likelh_pdf.pro
\startdata
%        G1.5     G1.3     G0.8     R50      R25     I50
G1.5  & 0.0    & 21.8    & 388.5  & 15.4   & 11.1   & 18.1   \\
G1.3  &\nodata & 0.0     & 154.1  & 2.3    & 2.2    & 2.3    \\
G0.8  &\nodata & \nodata & 0.0    & 209.1  & 159.8  & 218.7  \\
R50   &\nodata & \nodata &\nodata &  0.0   & 0.2    & 0.2    \\
R25   &\nodata & \nodata &\nodata &\nodata &  0.0   & 0.1    \\
I50   &\nodata & \nodata &\nodata &\nodata &\nodata & 0.0    \\
\enddata
\tablecomments{Assumes mock SDSS DR7 data set of 1500 quasars at $z=2.5$, with $\snr=4$
in the \lya\ forest.}
\end{deluxetable}

We find that the PDF can distinguish G0.8 from G1.5 with a high significance of
$\lnl({\mathrm G0.8|G1.5}) =388.5$. At first glance, we get the surprising result that the PDF
can differentiate the inhomogeneous \heii\
 reionization model R50 from the fiducial
G1.5 at a significant $\lnl({\mathrm R50|G1.5})= 15.4$. However, note
that the models R50 and G1.3 are
approximately degenerate with $\lnl({\mathrm R50|G1.3})=2.3$ 
between the two models. 
This is because the PDF of R50 is 
essentially averaged across its two different phases of $\gamma=1.2$ and 
$\gamma=1.5$, thus it can be fit by a homogeneous model with
 some intermediate value of $\gamma$.

As expected, we find that the PDF is degenerate between the models R50, R25, and I50
since the only difference is the spatial distribution of the line-of-sight 
segments which
intersect with hot bubbles. 

There are several uncertainties in the IGM parameters which might 
be degenerate with $\gamma$ in the flux PDF. The Jean's smoothing scale of the 
IGM is dependent on physics such as complex hydrodynamic effects and the
temperature evolution of the gas, which are not very well understood.
\citet{gned+hui98} have argued for an effective smoothing scale which is 
approximately half the Jean's scale at the epoch of observation. This
gives $\sigma_{eff} \approx 0.15 \mpc$ at $z=2$. As the \citet{slosar+09} 
simulations used in this paper do not accurately capture small-scale power
($\sigma_{eff} = 0.25 \mpc$ was used), we turn to another set of simulations:
the \citet{white+10} `Roadrunner' simulations are similar to those 
of \citet{slosar+09} 
except with higher resolution ($0.1875 \mpc$ grid size vs $0.5 \mpc$) and 
smaller box ($(750 \mpc)^3$ vs $(1500\mpc)^3$).
The smoothing scale used in the Roadrunner simulations is  
$\sigma_{eff}=0.1 \mpc$.
To approximate the uncertain pressure smoothing scale 
on the flux PDF, 
we take the optical depth outputs of these simulations and smooth them
to an overall smoothing scale of $\sigma_{eff}=0.15 \mpc$ using a 
Gaussian kernel. In comparison with the original spectra, we find that the
resulting flux PDFs differ by only 
$\lnl(\sigma_{eff}=0.15\mpc|\sigma_{eff}=0.1\mpc) \approx 2$, 
which is significantly 
less than the difference caused by varying $\gamma$. 
In any case, in the future data analysis we expect to use hydrodynamic 
simulations which would obviate the need to explicitly assume a value
for the Jean's smoothing scale. 

Another systematic which could be degenerate with $\gamma$ are the 
uncertainties in the value of the mean flux of the \lya\ forest \fmean, 
which we have thus far assumed to
be fixed. Using the somewhat smaller SDSS Data Release 5, \citet{dall+09} 
reported errors in their measurements of $\tau_{eff}\equiv -\ln \fmean$ 
equivalent to $\sigma_F \approx 0.3 \%$ in the mean flux at $z \approx 2.5$.
We study the effect of this uncertainty by assuming the actual
mean flux is distributed as a Gaussian distribution
with $\sigma_F=0.3 \%$, and marginalizing over this 
when calculating the likelihoods between the different values of $\gamma$.
This gives a value of $\lnl(G1.3|G1.5) \approx 15$ in comparison with 
 $\lnl(G1.3|G1.5,\delta\fmean=0) \approx 20$ reported in 
Table~\ref{tab:pdflike}
This is a significant effect and needs to be taken into account
in a more formal data analysis, but 
does not qualitatively affect our conclusions.

In general, we have found that the flux PDF from the latest SDSS data 
can be used to constrain the equation of state of the IGM 
if a homogeneous IGM is assumed. Indeed, the evolution of $\gamma$ 
with redshift can potentially be measured.
In the SDSS DR7 quasar catalog there are more than 
700 quasars with $\snr \geq 4$ in the \lya\ forest 
 within redshift bins of $\Delta z = 0.3$ up to $z_{qso} \sim 3.5$.
The logarithmic likelihood \lnl is roughly proportional to sample size, thus
in comparison with our mock sample size of 1500 quasars at $2.4 \leq z_{qso} \leq 2.7$
and the results of Table~\ref{tab:pdflike},
we expect to be able 
to measure the IGM equation of state in these redshift bins
to a precision of $\Delta \gamma \sim 0.2$
with a confidence of $\lnl \sim 10$ (approximately $4\sigma$), 
although it becomes more difficult to 
estimate accurately the quasar continuum in the \lya\ forest at higher redshifts.

These simulations suggest that an analysis on the flux PDF from the SDSS 
sample could detect the predicted 
 suppression in the equation of state from $\gamma \approx 1.5$ to $\gamma \lesssim 1$
\citep[see][]{furl+oh08,mcquinn+09}
due to \heii\ reionization at $z \sim 3$. 
Note that this approach is complementary to studies which used the evolution of 
the mean optical depth $\tau_\mathrm{eff}$ in the \lya\ forest to study \heii\ reionization, 
since the PDF can in principle be used to
measure $\gamma$ at fixed $\langle F \rangle = \exp({-\tau_\mathrm{eff}})$.

\section{Threshold Probability Functions from Mock Spectra}
\label{sec:stwo}

In this section, we evaluate the threshold correlation statistics \stwo, \ctwo,
and \dtwo\ on the mock \lya\ forest sample described in the previous sections.
We first describe the form of these functions, before applying them to the
various IGM models and investigate their ability to distinguish between models.

\subsection{Basic Form of \stwo, \ctwo, and \dtwo\ on the \lya\ Forest}
As we are primarily interested in breaking the degeneracy between the equation of state of the IGM 
and thermal inhomogeneities with comoving scales of $\sim 10 \mpc$,
we first smooth the 
mock \lya\ forest spectra with a Gaussian window of width $\sigma = 10 \mpc$. This
has the effect of increasing the contrast from the temperature inhomogeneities
and smoothing over the noise, although after smoothing the shapes of the
spectra are still dominated by instrumental noise and the large-scale structure of matter.
In other words, when visually inspecting the smoothed spectra of the same 
line-of-sight modified to the different IGM models, 
it is still impossible to tell {\it a priori} which is the inhomogeneous model.   

For each value of \fth,  we first identify the pixels which have $F \geq \fth$
within each individual smoothed spectrum, keeping track of the `clusters' 
of adjoined pixels.
$\stwo(r,\fth)$ is then calculated by counting pairs of pixels above
\fth, and separated by correlation length $r$.
This is then normalized to give a probability. We also keep track of 
$\ctwo(r,\fth)$ as the probability of pixel pairs which are within the same cluster
of pixels with $F \geq \fth$.
The threshold probability functions are then evaluated on a 2-dimensional grid in 
\fth\ and $r$ on the SDSS DR7 mock samples.

\begin{figure*}
\plottwo{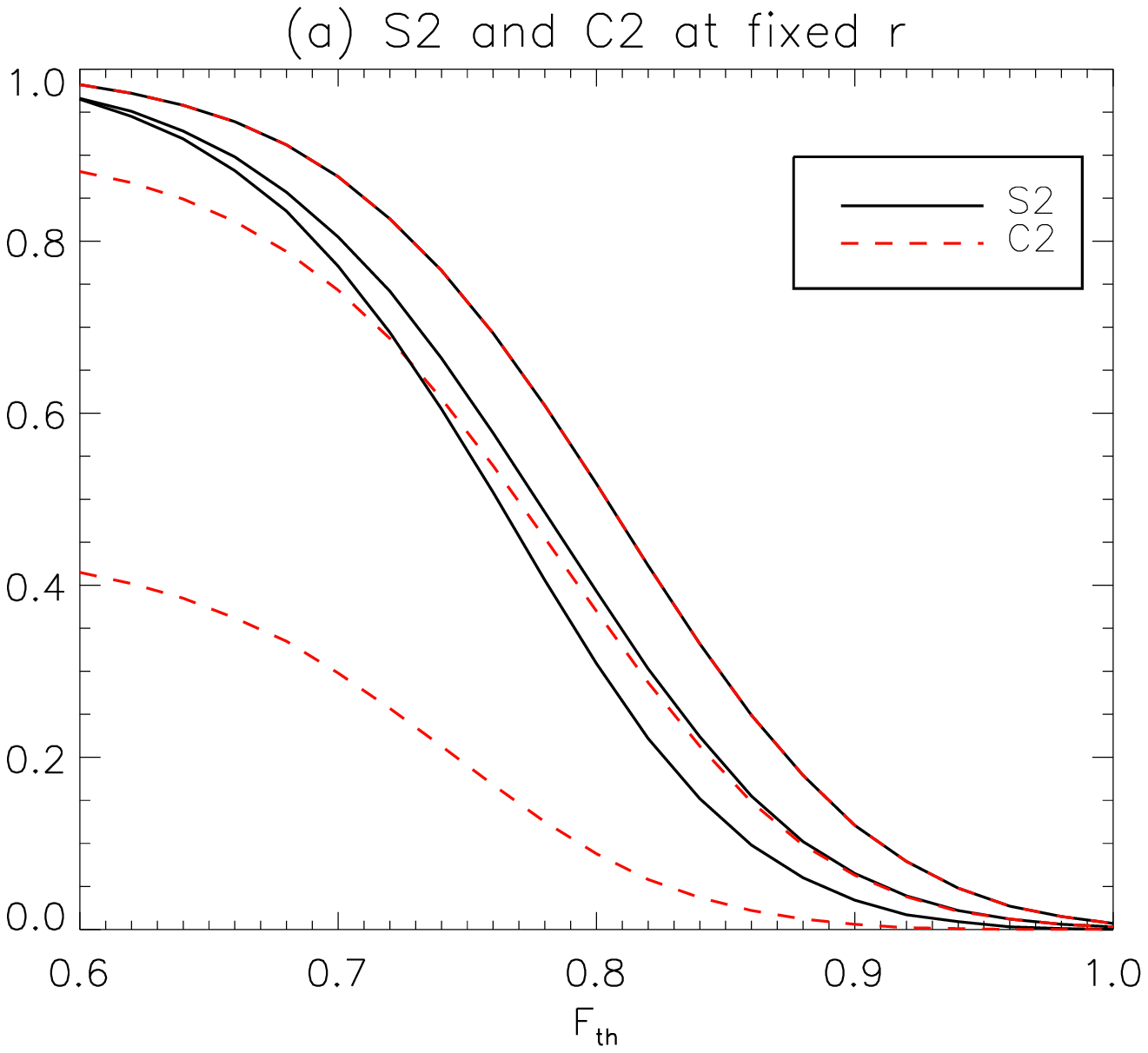}{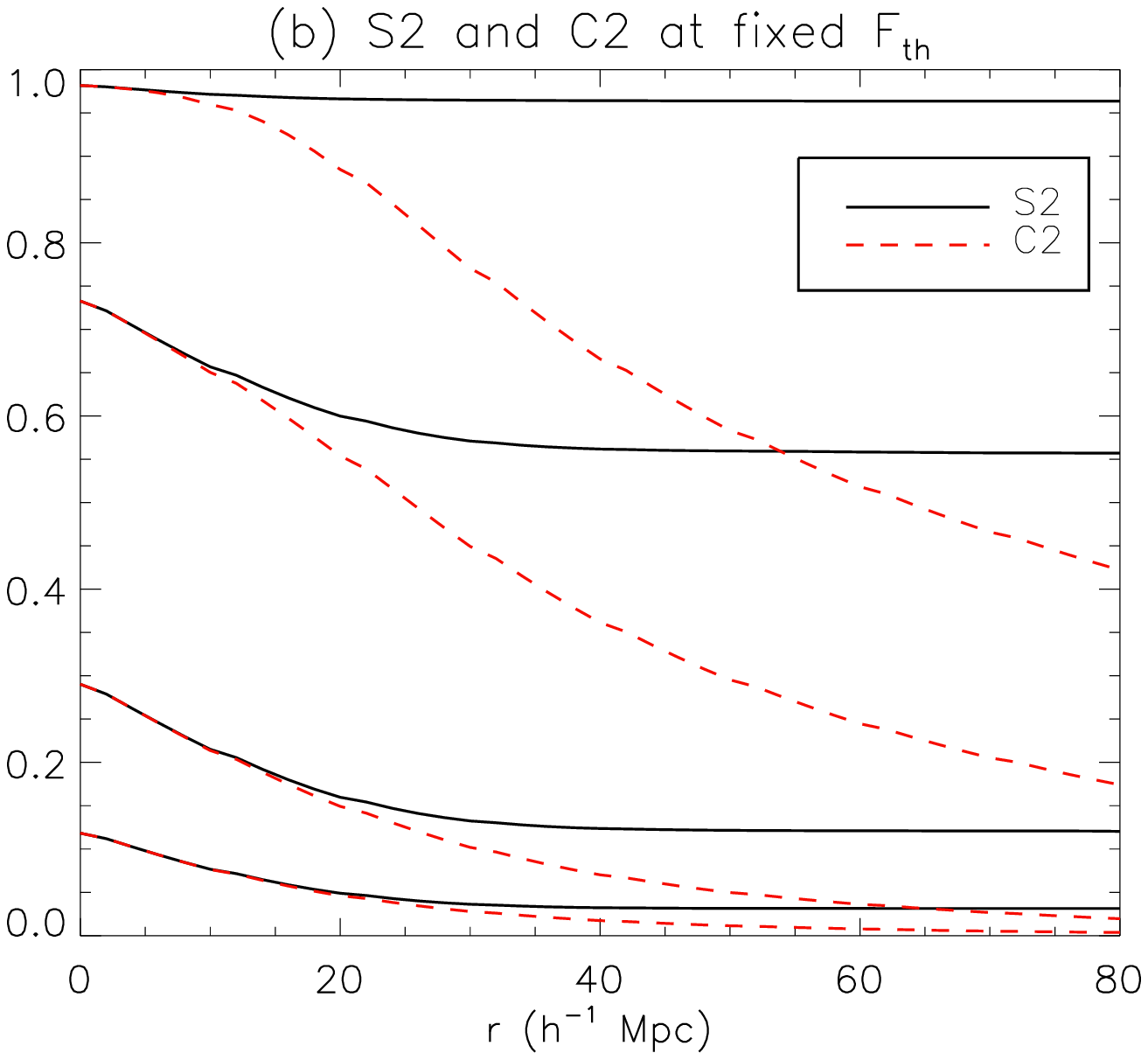}
\caption{\label{fig:s2_basic} (a) \stwo\ (solid black lines) and \ctwo\ (dashed 
red lines) from model G1.5 plotted as a function of flux threshold \fth\ at fixed comoving distances  
(from top to bottom) $r=0 \mpc, 15\mpc, 60\mpc$. Note that \stwo\ and \ctwo\ overlap 
in the case of $r=0\mpc$ (uppermost plot).
(b) \stwo\ and \ctwo\ plotted as a function of comoving distance $r$ at fixed
flux threshold $\fth=0.6, 0.75, 0.85, 0.9$ from top to bottom. See electronic edition of the Journal for a color version of this figure.
}
\end{figure*}

\begin{figure}
\epsscale{1.15}
% Generated from 'plotsurf' in tau_toy/2dcomp/plot_s2arr.pro 
\plotone{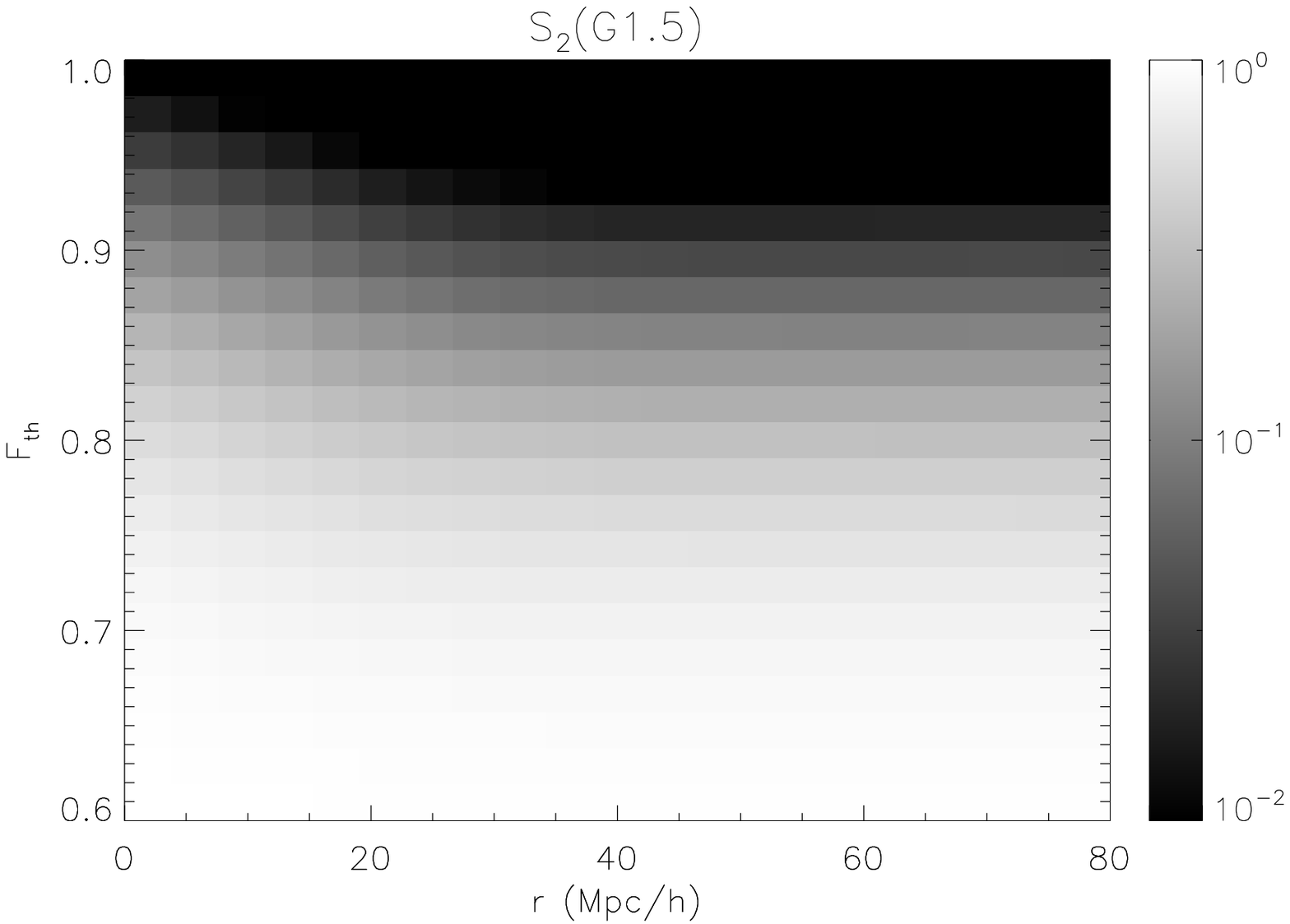} 
\caption{\label{fig:s2-2d} 2-dimensional density plot of \stwo\ as a function 
of $F$ and $r$, evaluated for one mock sample based on the model G1.5. }
\end{figure}

 Figure~\ref{fig:s2_basic}a plots 
$\stwo(\fth)$ and $\ctwo(\fth)$ at fixed $r$ for the fiducial model G1.5
 (recall that  $\stwo = \ctwo + \dtwo$, 
Equation~\ref{eq:stwodef}).
At large \fth\ few pixels in the smoothed spectra have
sufficiently large flux $F$ to rise above the threshold, thus
$\stwo(\fth | r)$ tends to zero at all $r$ as $\fth \rightarrow 1$.
Conversely, as the threshold is lowered below the mean flux $\langle F \rangle$, 
increasing numbers of pixels satisfy the criterion $F \geq \fth$
and \stwo\ trends towards unity, 
\be
\lim_{\fth  < \fbar} \stwo(\fth  | r)= 1.
\ee 

At small pixel separations $r$, the main contribution to \stwo\ comes
from \ctwo\ as most pixel pairings that rise above the flux threshold are within 
the same cluster of pixels (recall that `clusters' in this context refers to groupings
of contiguous pixels and not galaxy or stellar clusters). 
Note that $\stwo(\fth | r=0) \equiv \ctwo(\fth | r=0) $ is just the integral 
of the flux PDF (see Equation~\ref{eq:s20}).
The contribution of \ctwo\ to \stwo\ decreases and gives way to 
$\dtwo \equiv \stwo - \ctwo$  at greater $r$ as there is
greater probability of finding pixels from different clusters than within the same cluster.

Figure~\ref{fig:s2_basic}b plots \stwo\ and \ctwo\ as a function of $r$ at fixed values
of \fth. This shows more clearly the trend of \ctwo\ dominating the probability of 
pixels pairs being above the flux threshold at small separations. The overall
probability \stwo\ increases as the flux threshold is lowered and more pixels satisfy
the condition $F \geq \fth$. At separations larger than $r \gtrsim 50 \mpc$, 
\dtwo\ dominates and we see the asymptotic behaviour 
$\stwo(r | \fth) \rightarrow \stwo^2(0 | \fth)$ as the pixel separations at large scales
is effectively random. 
As $\stwo(r=0 | \fth)$ and its square essentially measure the integral
of the flux PDF (Equation~\ref{eq:s20}), we expect any additional information
from \stwo\ to come at small correlation lengths $r \lesssim 50$, i.e.\ from \ctwo.

Figure~\ref{fig:s2-2d} displays \stwo\ as a density plot in the two dimensions
of $r$ and \fth.

\subsection{Distinguishing Between IGM Models}

We compute \stwo\ on realizations of DR7 mock samples computed for 
the various IGM models
introduced in Section~\ref{sec:model}, using the systematics and mock sample previously
discussed (1500 lines-of-sight with \snr=4, flux errors $\sigma_F = 9\%$). 
The mock data are evaluated on 2-dimensional grids in 
the ranges $0 \mpc \geq r \geq 80 \mpc$ and $0.5 \geq \fth \geq 1.0$, with 26 bins 
in each dimension.

As was done for the PDFs in Section~\ref{subsec:pdf}, we use the logarithmic 
likelihood \lnl\ (Equation~\ref{eq:like}) to quantify the ability to differentiate
the different IGM models. From the initial 676 data points for $\stwo(r, \fth)$
from each model,  we first remove $\sim 20$ points with $\stwo < 10^{-3}$ to reduce 
the dynamic range,
and then remove a few more points in order to ensure that the covariance matrix 
(calculated from $\sim 100$ mock realizations for each model) is well-conditioned.

\begin{figure*}[t]
%\epsscale{1.15}
\plottwo{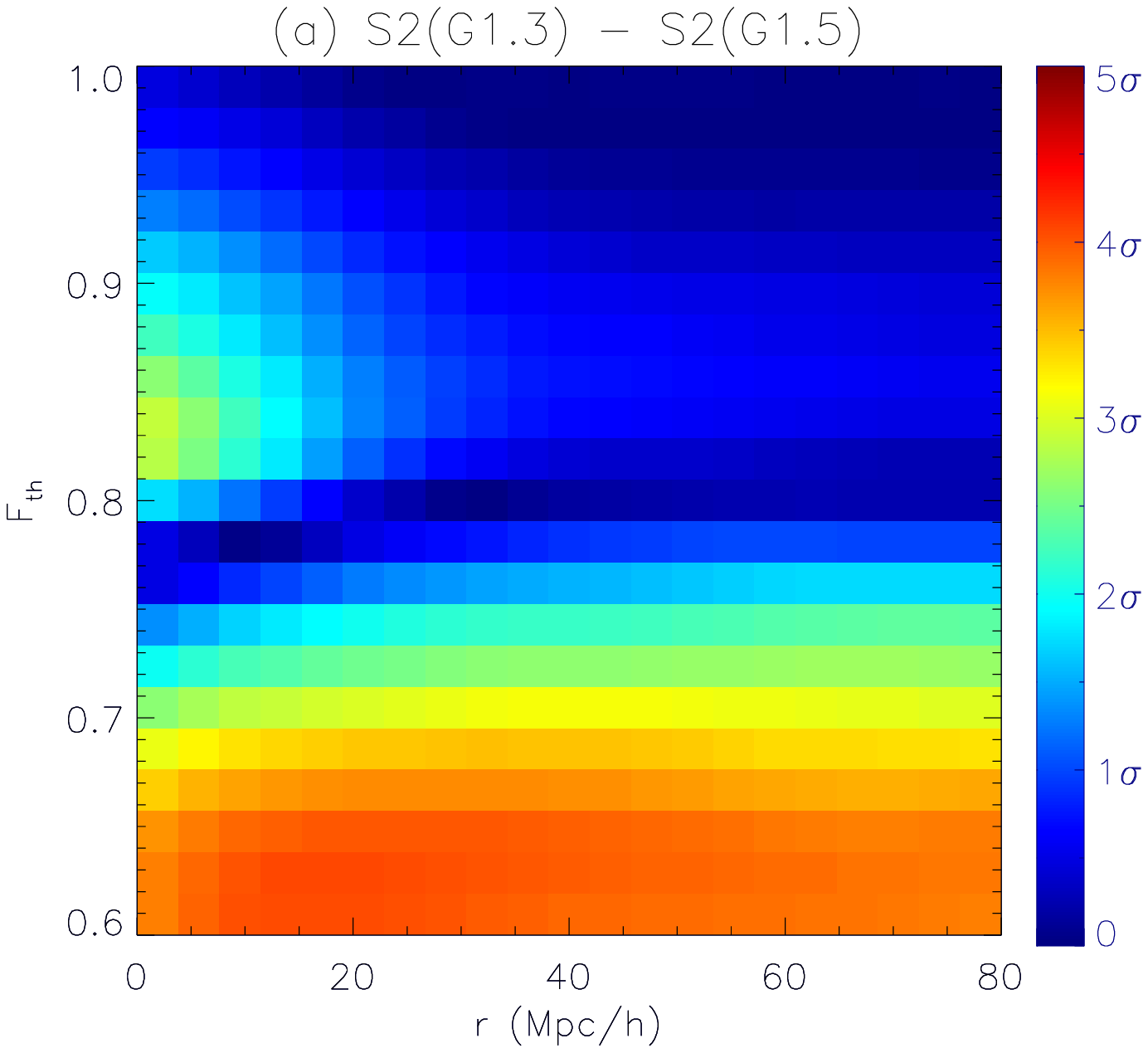}{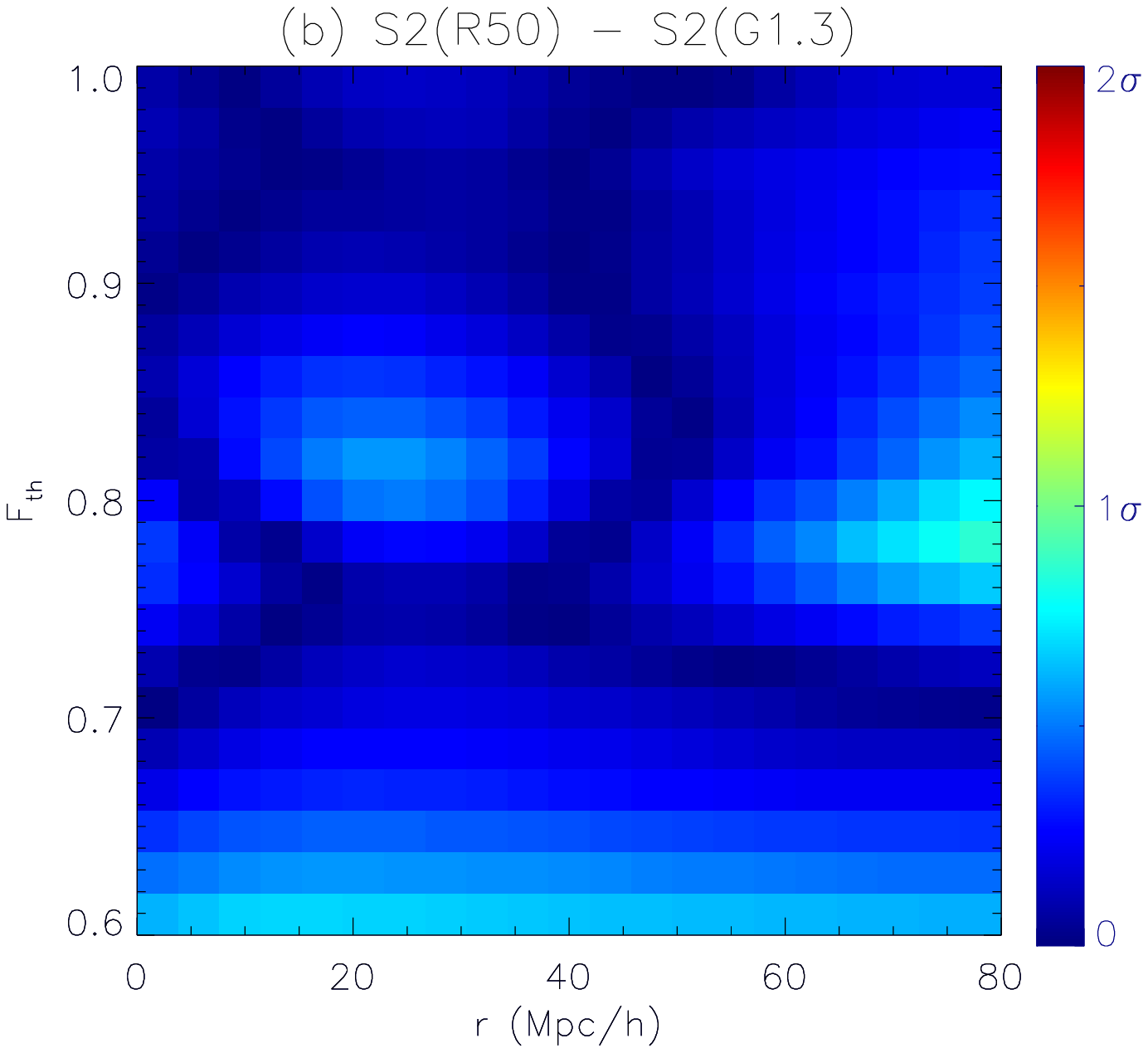}
\caption{\label{fig:2d_diff_fs9} (a) Absolute difference between \stwo\ 
from models G1.3 and G1.5 as a function of $r$ and \fth, 
normalized by the error from the former. (b) Absolute difference between 
\stwo\ from models R50 and G1.3. Note different intensity scale from plot on the
left. See electronic edition of the Journal for a color version of this figure.
}
\end{figure*}
\begin{figure*}
%\epsscale{1.15}
\plottwo{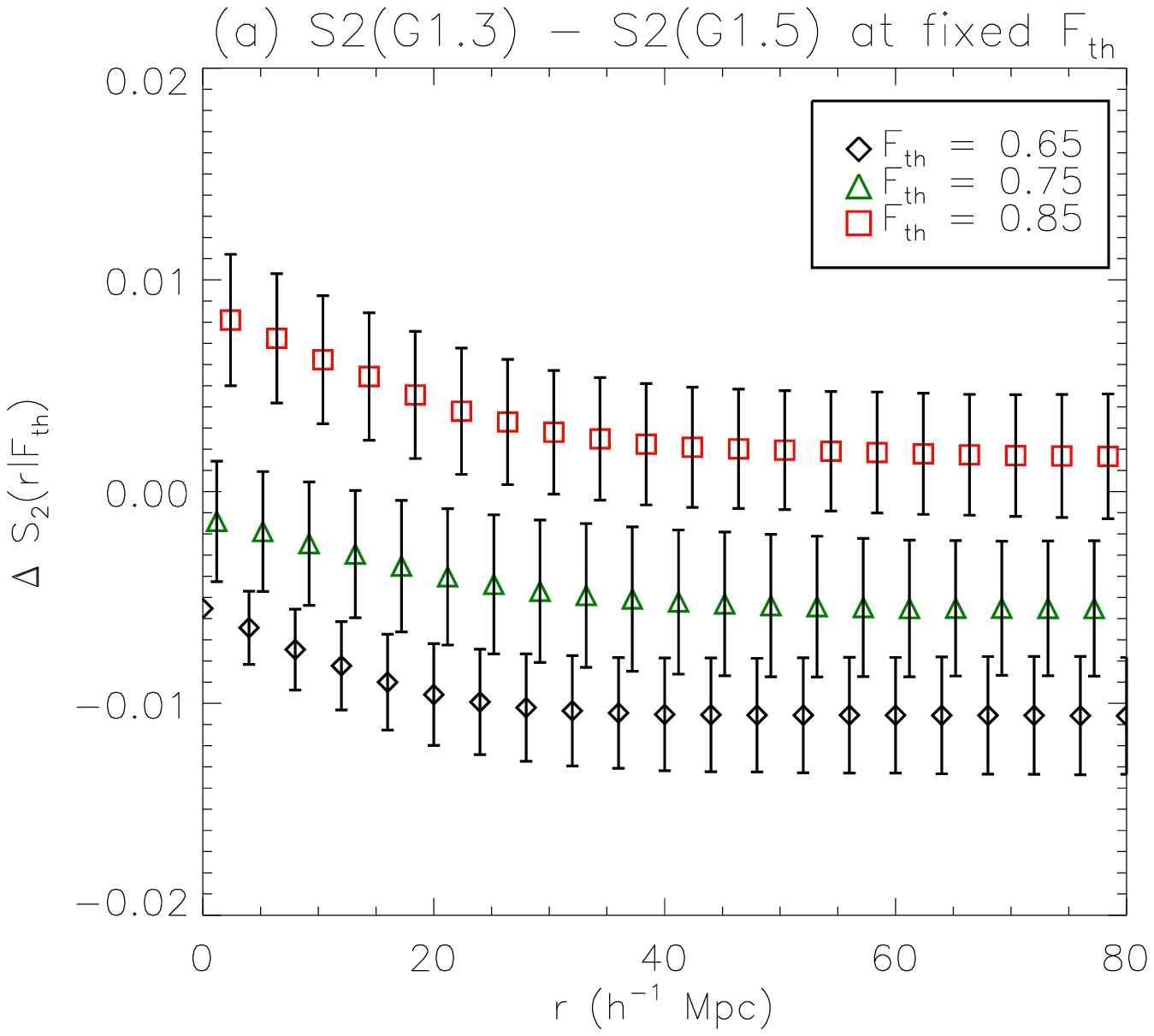}{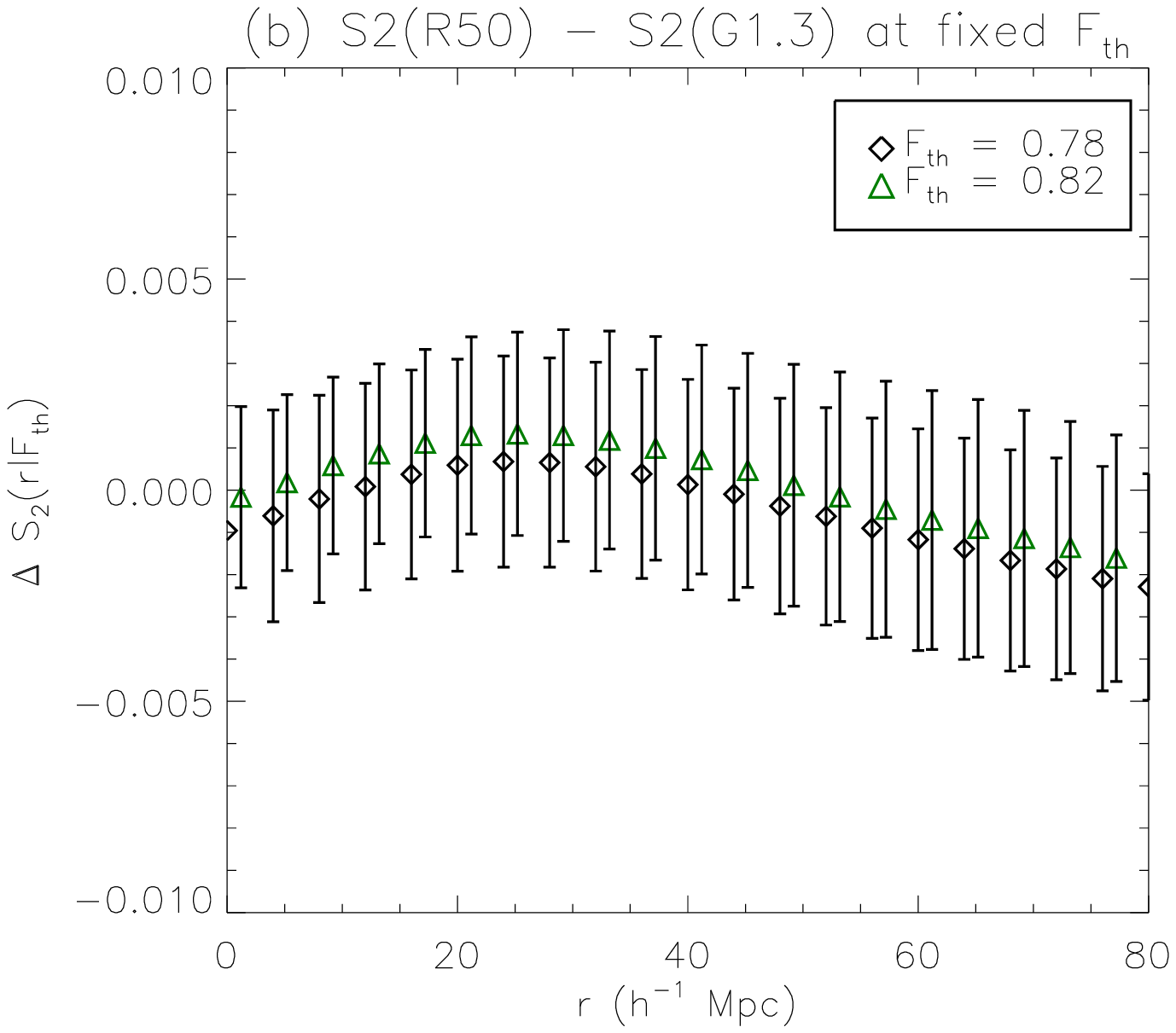}
\caption{\label{fig:diffr_fs9} (a) Difference between \stwo\ from models G1.3 and G1.5, 
plotted as a function of correlation length $r$ at fixed values of $\fth=0.65$ 
(black diamonds), $\fth=0.75$ (green triangles), and $\fth=0.85$ (blue squares). 
The points are offset horizontally for clarity. (b) Difference between \stwo\ from 
models R50 and G1.3, as a function of $r$ at fixed values of $\fth=0.78$ 
(black diamonds) and $\fth=0.82$ (green triangles). 
The different shape of (b) compared to (a) indicates the biasing of \stwo\ 
towards larger $r$ due to the presence of temperature inhomogeneities in R50.
See electronic edition of the Journal for a color version of this figure.}
\end{figure*}

The values of \lnl\ between the various models are
 summarized in Table~\ref{tab:s2like_fs9}. 
In general, we see that the threshold probability function does a somewhat better
job of distinguishing between the homogeneous models: $\lnl(G1.3|G1.5) = 54.1$ between
models G1.5 and G1.3 compared with $\lnl(G1.3|G1.5) = 21.8$ when 
using the PDF computed for the 
same mock samples (Table~\ref{tab:pdflike}).

The absolute differences \dstwo\ between the two models as a function of $r$ and \fth\ 
are shown in Figure~\ref{fig:2d_diff_fs9}a, normalized by the estimated error in each point.
Figure~\ref{fig:diffr_fs9}a plots the difference between these two models as a function 
of $r$ at several fixed values of \fth.
It can be seen that the plots of $\dstwo(r | \fth)$ have the same general shape as 
$\stwo(r | \fth)$ (Figure~\ref{fig:s2_basic}b) at the various values of \fth\ shown, 
which shows that \dstwo\ between these two models 
includes contributions from all comoving scales.

\begin{deluxetable}{l|cccccc}
\tablewidth{0pt}
\tablecaption{\label{tab:s2like_fs9} \lnl between $\stwo(r,\fth)$ for different IGM models, assuming 9\% flux continuum error}
\tablehead{ & \colhead{G1.5} & \colhead{G1.3} &\colhead{G0.8} &\colhead{R50} &\colhead{R25} &\colhead{I50} }
% Computed in 'pdflike' in tau_toy/pdf_cosvar/likelh_pdf.pro
\startdata
%        G1.5     G1.3     G0.8     R50      R25     I50
G1.5  & 0.0    & 54.1    & 837.6  & 67.8   & 65.3   & 69.1   \\
G1.3  &\nodata & 0.0     & 400.6  & 14.1   & 7.2    & 17.7    \\
G0.8  &\nodata & \nodata & 0.0    & 439.1  & 433.3  & 389.2  \\
R50   &\nodata & \nodata &\nodata &  0.0   & 4.5    & 3.2    \\
R25   &\nodata & \nodata &\nodata &\nodata &  0.0   & 4.1    \\
I50   &\nodata & \nodata &\nodata &\nodata &\nodata & 0.0    \\
\enddata
\tablecomments{Assumes mock SDSS DR7 data set of 1500 quasars at $z=2.5$, with $\snr=4$
in the \lya\ forest.}
\end{deluxetable}

However, \stwo\ has some ability to
break the degeneracy between the models G1.3 and R50 with $\lnl(R50|G1.3)=14.1$, 
compared with $\lnl(R50|G1.3)=2.3$ using the PDF. Figure~\ref{fig:2d_diff_fs9}b shows 
$| \dstwo(r,\fth)|$ for models R50 and G1.3,
 and we see that the differences are relatively subtle with
$| \dstwo(r,\fth)| \lesssim 1 \sigma$. In Figure~\ref{fig:diffr_fs9}b,
the plots of $\dstwo(r | \fth)$  shows that the deviations do not trace
the shape of $\stwo(r | \fth)$, but have a hump-like shape
peaking at scales of $r \sim 20 \mpc$ and extending well into the larger scales where
 $\stwo(r | F_{th})$ goes flat in Figure~\ref{fig:s2_basic}b . 
A comparison with Figure~\ref{fig:s2_basic}b suggests that most of this contribution 
comes from \ctwo.
However, if the bubbles have a smaller characteristic size then they are harder to 
distinguish from the closest homogeneous model:
 $\lnl(R25|G1.3) = 7.2$ which would be a marginal detection. 

As we have discussed, the temperature inhomogeneities in the IGM are too subtle to 
overcome the density field which dominates the optical depth distribution of the IGM, 
but regions of higher temperature 
such as those shown in Figure~\ref{fig:fluxplot} can broaden the width of 
pixel clusters
that rise above the flux threshold \fth\ when 
averaged across many spectra, increasing \stwo\ at the scales represented by the 
various path lengths at which the \lya\ lines-of-sight intersect the hot bubbles.
The shape of \dstwo\ in Figure~\ref{fig:s2_basic}b represents a slight broadening
of \ctwo\ from the hot bubbles in the model.
While a homogeneous IGM with $\gamma \approx 1.3$ can provide a flux PDF that
is indistinguishable from an inhomogeneous model, the threshold 
correlation functions measures sufficient spatial information to detect
the temperature inhomogeneities and break the degeneracy with the 
best-fit homogeneous model.

In addition, we see from Table~\ref{tab:s2like_fs9} that the model R25 with 
smaller hot bubbles is difficult to distinguish from model R50 with 
$\lnl(R25|R50) = 4.5$, but the fact that \lnl\ is not of order unity indicates
 that \stwo\ encodes some information
on the characteristic scale of the temperature inhomogeneities. 
Looking at the inverted bubble model I50, we see that it is nearly degenerate with
R50, with $\lnl=3.2$ between the two models.
Nevertheless, it would appear that all the inhomogeneous models can 
be differentiated from G1.3 with $\lnl \gtrsim 10$ in comparison with $\lnl \sim 1$
when using the PDF.

\subsection{Improved Flux Continuum Estimates}

In our mock spectra we have so far assumed what we regard as a worst-case scenario
for estimating the flux continuum of the \lya\ forest, with errors of around 9\%.
This was the uncertainty that \citet{suz+05} found from PCA fitting on the intrinsic
quasar spectrum redwards of the \lya\ emission line and without using any information 
from the \lya\ forest itself. 

There are various possibilities to improve on the flux continuum fits. One method is to assume
that the mean flux \fbar\ for each \lya\ forest 
is equal to the global mean flux $\langle F(z) \rangle$
at the corresponding redshift, and normalize the observed quasar continuum to this value
 (N. Suzuki, private communication). 
In this way, the errors in the continuum fit 
would be limited to a combination of cosmic variance and errors in the determination
of the mean flux. 
The dispersion in the mean flux between different lines-of-sight is of order 1-2\%,
while within an individual line-of-sight with e.g.\ $\snr \sim 4$ across $\sim 500$ pixels 
the mean flux can be determined  to about $(S/N \times \sqrt(500))^{-1} \sim 1 \%$.
In principle, this yields an error on the continuum
determination at the few percent level (N. Suzuki, private communication).

In this subsection, we compute the threshold probability functions for 
mock spectra with assumed flux continuum errors of
$\sigma_F = 3\%$, which is an optimistic estimate of the precision believed possible
with the mean flux fitting method. In all other respects, the properties of our 
mock sample are unchanged from the previous sections. 

\begin{deluxetable}{l|cccccc}
\tablewidth{0pt}
\tablecaption{\label{tab:s2like_fs2} \lnl between $\stwo(r,\fth)$ for different IGM models, assuming 3\% flux continuum error}
\tablehead{ & \colhead{G1.5} & \colhead{G1.3} &\colhead{G0.8} &\colhead{R50} &\colhead{R25} &\colhead{I50} }
% Computed in 'pdflike' in tau_toy/pdf_cosvar/likelh_pdf.pro
\startdata
%        G1.5     G1.3     G0.8     R50      R25     I50
G1.5  & 0.0    & 150.1   &1620.7  & 205.3  & 181.0  & 191.1  \\
G1.3  &\nodata & 0.0     & 779.6  & 29.0   & 21.9   & 28.6   \\
G0.8  &\nodata & \nodata & 0.0    & 756.8  & 707.2  & 965.9  \\
R50   &\nodata & \nodata &\nodata &  0.0   & 15.1   & 2.7    \\
R25   &\nodata & \nodata &\nodata &\nodata &  0.0   & 10.4   \\
I50   &\nodata & \nodata &\nodata &\nodata &\nodata & 0.0    \\
\enddata
\tablecomments{Assumes mock SDSS DR7 data set of 1500 quasars at $z=2.5$, with $\snr=4$
in the \lya\ forest.}
\end{deluxetable}

\begin{figure*}
%\epsscale{1.15}
\plottwo{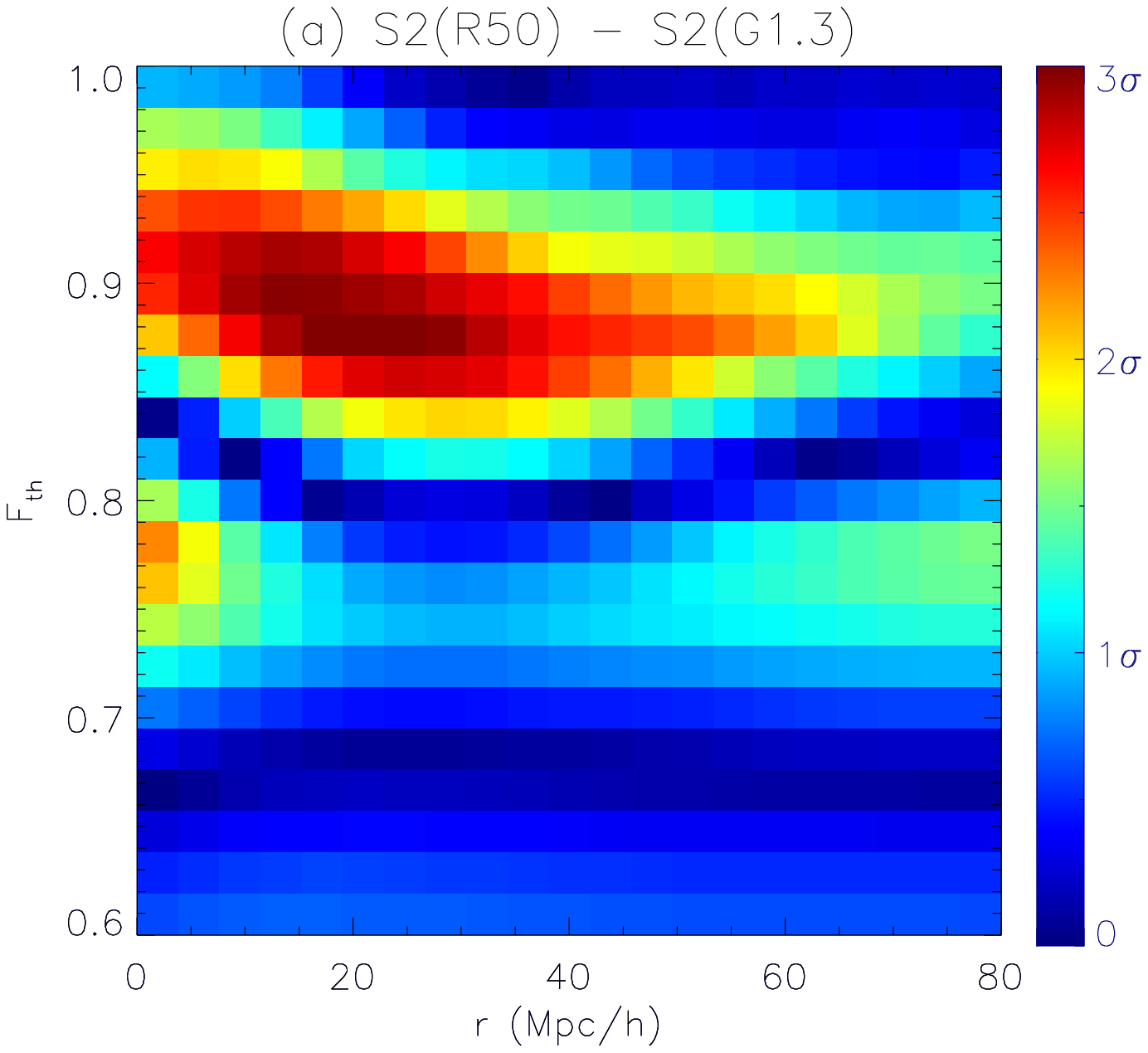}{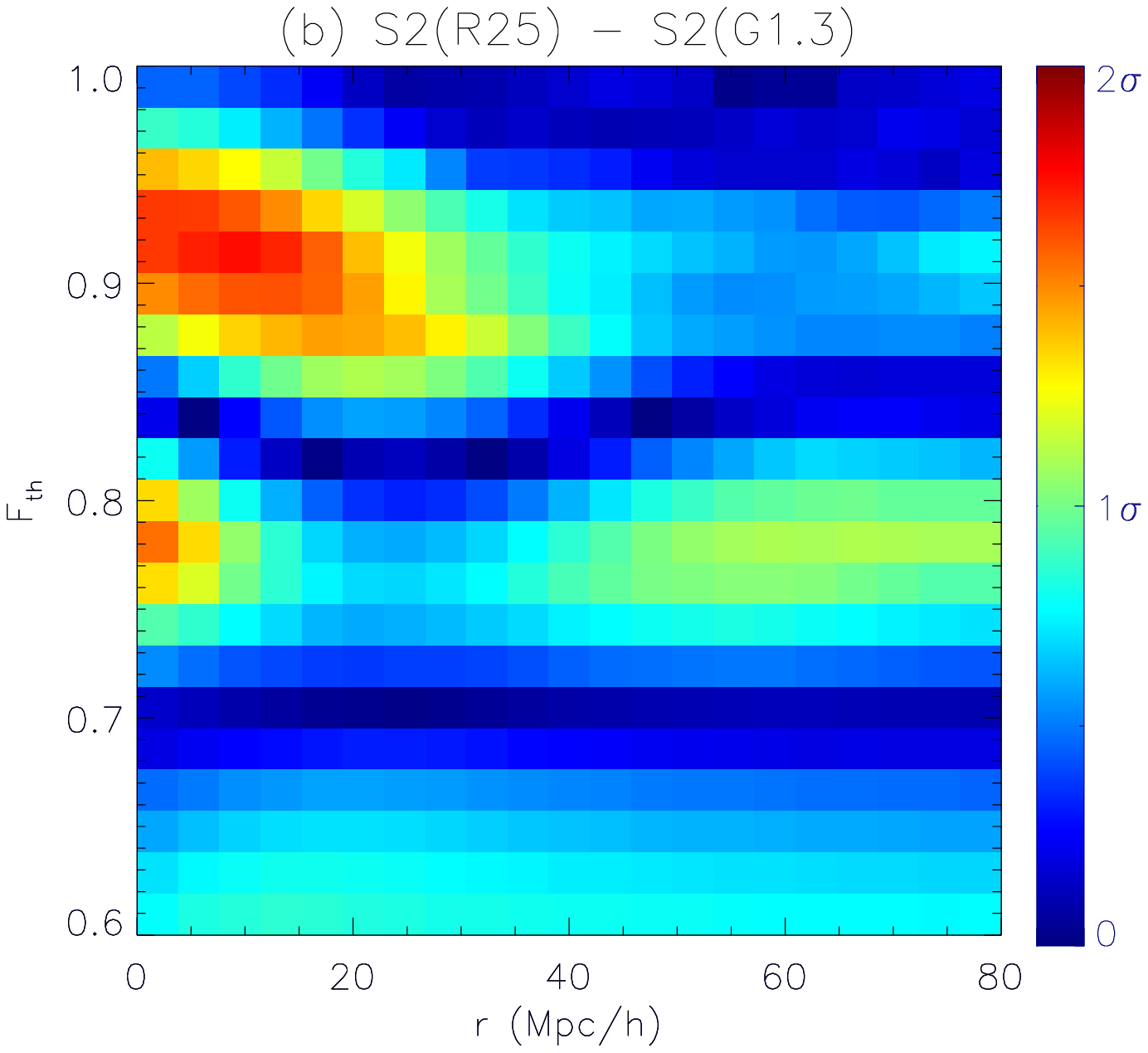}
\caption{\label{fig:2d_diff_fs2} (a) Absolute difference between \stwo\ 
from models R50 and G1.3 as a function of $r$ and \fth, 
normalized by the error from the former. These were calculated assuming
3\% errors in the continuum determination. (b) Absolute difference between 
\stwo\ from models R25 and G1.3, assuming 3\% continuum errors. Note that the differences in (b) peak
at smaller scales.
}
\end{figure*}

Table~\ref{tab:s2like_fs2} summarizes the logarithmic likelihoods for differentiating
\stwo\ calculated from any two IGM models. Overall, the values of \lnl see
 a marked improvement
when compared with the values in Table~\ref{tab:s2like_fs9} computed assuming 9\% flux
continuum errors. 
It is now possible to tell the inhomogeneous models apart from the model G1.3 with 
significant confidence, with $\lnl \gtrsim 30$.

\begin{figure}
%\epsscale{1.15}
\plotone{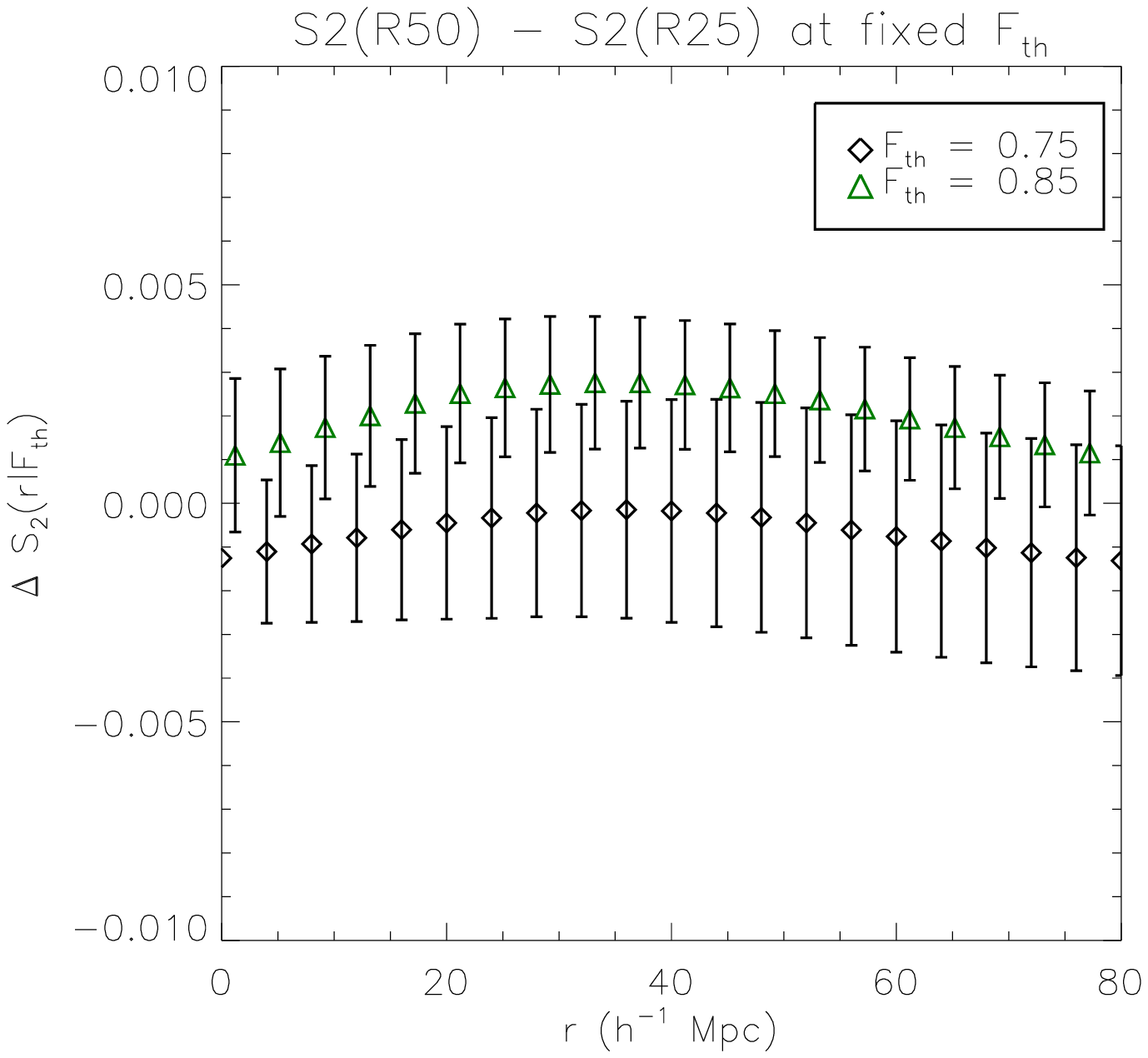}
\caption{\label{fig:diffr_R100_fs2} Difference between \stwo\ from models R50 and R25, 
plotted as a function of correlation length $r$ at fixed values of $\fth=0.75$ 
(black diamonds) and $\fth=0.85$ (green triangles). 
The points are offset horizontally for clarity. 
See electronic edition of the Journal for a color version of this figure.
}
\end{figure}

More importantly, \stwo\ can now place significant constraints on the characteristic
scale of the \heii\ reionization bubbles, as $\lnl(R25|R50)=15.1$. 
The differences can be clearly seen in the two-dimensional differences plots
of $\stwo(r, \fth)$ of these models with
respect to that calculated from the best-fit homogeneous model G1.3,
Figure~\ref{fig:2d_diff_fs2}. The deviations from the model G1.3 peak
at smaller scales $r$ in the case of R25, Figure~\ref{fig:2d_diff_fs2}b, 
compare R50,  Figure~\ref{fig:2d_diff_fs2}a.
This effect can be seen more emphatically in Figure~\ref{fig:diffr_R100_fs2},
which shows \dstwo\ between R25 and R50 plotted at several flux threshold values.
At the flux threshold values shown, R50 displays a distinct
increase in \stwo\ at larger scales compared with R25. 

Even with the improved precision for the flux continuum,  \stwo\ is still unable
to distinguish between R50 and its topologically inverted counterpart I50, with 
$\lnl = 2.7$ for differentiating the two models. 
This is not surprising, as Figure~\ref{fig:bubbles} shows that the size distribution
of hot and cold regions in the R50 and I50 boxes are similar due to the equal 
volume fractions. 

\section{Discussion \& Conclusion}

In this paper we have introduced to astrophysics a set of new correlation statistics,
\stwo,
\ctwo, and \dtwo, that are evaluated as two-dimensional functions of
correlation length $r$ and transmitted flux threshold \fth. 
These `threshold probability functions' were tested on mock \lya\ forest spectra in which
instrumental noise and systematics were added at a level appropriate 
to the SDSS \lya\ forest
data, and we have made assumptions on the flux continuum errors based on the PCA
fitting method of \citet{suz+05}. 

\subsection{Flux PDF}  

As the threshold correlation statistics can be thought of as the flux probability
distribution function (PDF) measured as a function of spatial scale, we first
 computed the flux PDF in order to check that the errors from our mock spectra
are comparable with those of the observed PDF from SDSS spectra published
in \citet{desj+07}. At the level of uncertainty arising from the latest SDSS data set, 
we find that with the assumption of a homogeneous IGM the flux PDF
can in fact constrain the IGM equation of state, $\gamma$, to 
$\Delta \gamma \sim 0.1$ at the $\sim 4 \sigma$ level in redshift bins of 
$\Delta z \sim 0.3$. This would allow the redshift evolution of $\gamma$ to be constrained
through the epoch of \heii\ reionization at $z \approx 3$.

To-date, the best constraints on the 
equation of state $\gamma$ of the IGM have come from flux PDFs measured from 
small numbers ($\sim 20$) of high-resolution
\lya\ forest spectra, which have favored an inverted equation of state $\gamma < 0$
\citep{becker+07,kim+07,viel+09}. 
In a subsequent paper we aim to measure the flux PDF from SDSS DR7, and
in conjunction with numerical simulations and detailed modelling of systematics
make independent measurements of $\gamma$ as a function of redshift.
This would be an interesting complement to the studies using the mean flux
evolution in the \lya\ forest, which have placed the completion of \heii\ reionization at
$z \approx 3.2$ \citep{bern+03}.

 Baryon Oscillation Spectroscopic Survey (BOSS), part of the third phase of
the SDSS (SDSS-III), aims to observe $\sim$ 150,000 quasar lines-of-sight. 
While many of these \lya\ forest spectra will be of low signal-to-noise
($\snr \sim 1$), there will be sufficient numbers of moderate \snr\ 
spectra that even
better constraints can be made on the \heii\ end-of-reionization epoch than the DR7 sample 
considered in this paper.

\subsection{Threshold Probability Functions}

We have tested the threshold probability functions on the mock spectra, 
and found that they are able to break the degeneracy between an IGM with temperature
inhomogeneities 
and the best-fit homogeneous model of $\gamma=1.3$, 
at a confidence of $\sim 5 \sigma$.

If the errors in the continuum fitting can be reduced to $\sigma_F \approx 3\%$, 
then \stwo\ 
is sensitive to the characteristic scale of the temperature inhomogeneities, 
being able to distinguish between toy models with 50\mpc\ and 25\mpc\ bubbles
at $\sim 5 \sigma$. 

In this paper we have taken a simplified approach towards the instrumental systematics
and noise in the mock spectra. This has provided adequate estimates for our errors,
but would be insufficient to constrain the IGM from real data. 
Some of the systematics we have not taken into account include the possible influence of
damped \lya\ (DLA) regions in the \lya\ forest, metal line absorption, etc.
All these need to be modelled in detail when analyzing real data.
As for the flux continuum fitting, we hope to use methods such as mean-flux fitting
to reduce the errors
to several percent, which would enable the threshold probability functions to 
constrain the physical scale of any thermal inhomogeneities in addition to making
a detection.
Whatever the method used, the residual errors that arise need to be 
well-characterized in order to be modelled in the data analysis.

The mock spectra used in this paper have been generated using dark matter-only
simulations that do not capture detailed IGM physics, and various toy models for the 
IGM have been `painted' on to the basic set of mock spectra. In the actual data analysis
we anticipate fitting the data to mock spectra based on more detailed numerical
simulations that include the physics of \heii\ reionization.
 This would allow, in addition to a direct measurement of the temperature
range $\Delta T$ and spatial scale of the inhomogeneities, constraints to be placed on 
the underlying physical mechanisms such as the quasar luminosity function and duty cycle,
background ionization rate, gas clumping factors etc.

In the near future, the high area density of the BOSS quasars will enable correlation 
studies in the transverse direction between quasar pairs. 
It would be straightforward to extend the threshold probability functions to 
work in both the parallel and transverse directions relative to line-of-sight
 in order to utilize the full power of the BOSS data. 
In addition, transverse studies would ameliorate the effects of the uncertain 
fitting of the quasar continuum. 
However, we defer the three-dimensional generalization of the threshold probability functions 
 to a future paper.

\subsection{Summary}
Using mock \lya\ forest spectra based on toy models of the IGM and simulations
of the existing SDSS data, we have shown that detailed statistical analysis of these
spectra can provide insight into the physics of the IGM:

\begin{itemize}
\item The flux PDF from the SDSS DR7 can place significant constraints on the 
equation of state $\gamma$ of a homogeneous IGM to $\Delta \gamma \sim 0.1$ at
$z \approx 2.5$, and track its evolution through the end of \heii\ reionization. \\
\item We have introduced the threshold probability functions \stwo, \ctwo, and \dtwo,
which measure the \lya\ forest as functions of flux level and spatial scale, 
and have 
shown that they can differentiate an inhomogeneous IGM from the best-fit homogeneous
model at $\gtrsim 3 \sigma$.\\
\item If the flux continuum fitting can be carried out to $\approx 3\%$
accuracy, 
the threshold statistics can place constraints on the characteristic scale of the 
temperature inhomogeneities.
\end{itemize}

\acknowledgements{
The authors thank Sal Torquato, Matt McQuinn, Michael Strauss and Nao Suzuki for useful 
discussions and comments, 
and are grateful to Martin White for making available his Franklin 
and Roadrunner simulations. We also acknowledge the useful comments
 of the anonymous referee, which significantly improved this paper.
}

\bibliography{ms,apj-jour}

\begin{thebibliography}{41}
\expandafter\ifx\csname natexlab\endcsname\relax\def\natexlab#1{#1}\fi

\bibitem[{{Becker} {et~al.}(2007){Becker}, {Rauch}, \& {Sargent}}]{becker+07}
{Becker}, G.~D., {Rauch}, M., \& {Sargent}, W.~L.~W. 2007, \apj, 662, 72

\bibitem[{{Bernardi} {et~al.}(2003){Bernardi}, {Sheth}, {SubbaRao}, {Richards},
  {Burles}, {Connolly}, {Frieman}, {Nichol}, {Schaye}, {Schneider}, {Vanden
  Berk}, {York}, {Brinkmann}, \& {Lamb}}]{bern+03}
{Bernardi}, M., {Sheth}, R.~K., {SubbaRao}, M., {Richards}, G.~T., {Burles},
  S., {Connolly}, A.~J., {Frieman}, J., {Nichol}, R., 
  {et al.} 2003, \aj, 125, 32

\bibitem[{{Cen} {et~al.}(1994){Cen}, {Miralda-Escud{\'e}}, {Ostriker}, \&
  {Rauch}}]{cen+94}
{Cen}, R., {Miralda-Escud{\'e}}, J., {Ostriker}, J.~P., \& {Rauch}, M. 1994,
  \apjl, 437, L9

\bibitem[{{Croft} {et~al.}(2002){Croft}, {Weinberg}, {Bolte}, {Burles},
  {Hernquist}, {Katz}, {Kirkman}, \& {Tytler}}]{croft+02}
{Croft}, R.~A.~C., {Weinberg}, D.~H., {Bolte}, M., {Burles}, S., {Hernquist},
  L., {Katz}, N., {Kirkman}, D., \& {Tytler}, D. 2002, \apj, 581, 20

\bibitem[{{Croft} {et~al.}(1998){Croft}, {Weinberg}, {Katz}, \&
  {Hernquist}}]{croft+98}
{Croft}, R.~A.~C., {Weinberg}, D.~H., {Katz}, N., \& {Hernquist}, L. 1998,
  \apj, 495, 44

\bibitem[{{Dall'Aglio} {et~al.}(2009){Dall'Aglio}, {Wisotzki}, \&
  {Worseck}}]{dall+09}
{Dall'Aglio}, A., {Wisotzki}, L., \& {Worseck}, G. 2009, ArXiv e-prints

\bibitem[{{Dav{\'e}} {et~al.}(1999){Dav{\'e}}, {Hernquist}, {Katz}, \&
  {Weinberg}}]{dave+99}
{Dav{\'e}}, R., {Hernquist}, L., {Katz}, N., \& {Weinberg}, D.~H. 1999, \apj,
  511, 521

\bibitem[{{Desjacques} {et~al.}(2007){Desjacques}, {Nusser}, \&
  {Sheth}}]{desj+07}
{Desjacques}, V., {Nusser}, A., \& {Sheth}, R.~K. 2007, \mnras, 374, 206

\bibitem[{{Fan} {et~al.}(2002){Fan}, {Narayanan}, {Strauss}, {White}, {Becker},
  {Pentericci}, \& {Rix}}]{fan+02}
{Fan}, X., {Narayanan}, V.~K., {Strauss}, M.~A., {White}, R.~L., {Becker},
  R.~H., {Pentericci}, L., \& {Rix}, H. 2002, \aj, 123, 1247

\bibitem[{{Faucher-Gigu{\`e}re} {et~al.}(2008){Faucher-Gigu{\`e}re},
  {Prochaska}, {Lidz}, {Hernquist}, \& {Zaldarriaga}}]{fg+08}
{Faucher-Gigu{\`e}re}, C., {Prochaska}, J.~X., {Lidz}, A., {Hernquist}, L., \&
  {Zaldarriaga}, M. 2008, \apj, 681, 831

\bibitem[{{Furlanetto} \& {Oh}(2008{\natexlab{a}})}]{furl+oh08-2}
{Furlanetto}, S.~R. \& {Oh}, S.~P. 2008{\natexlab{a}}, \apj, 682, 14

\bibitem[{{Furlanetto} \& {Oh}(2008{\natexlab{b}})}]{furl+oh08}
---. 2008{\natexlab{b}}, \apj, 681, 1

\bibitem[{{Gnedin} \& {Hui}(1998{\natexlab{a}})}]{gned+hui97}
{Gnedin}, N.~Y. \& {Hui}, L. 1998{\natexlab{a}}, \mnras, 296, 44

\bibitem[{{Gnedin} \& {Hui}(1998{\natexlab{b}})}]{gned+hui98}
---. 1998{\natexlab{b}}, \mnras, 296, 44

\bibitem[{{Hui} \& {Gnedin}(1997)}]{hui+gned97}
{Hui}, L. \& {Gnedin}, N.~Y. 1997, \mnras, 292, 27

\bibitem[{{Jenkins} \& {Ostriker}(1991)}]{jen+ost91}
{Jenkins}, E.~B. \& {Ostriker}, J.~P. 1991, \apj, 376, 33

\bibitem[{Jiao {et~al.}(2009)Jiao, Stillinger, \& Torquato}]{jiao_torq09}
Jiao, Y., Stillinger, F.~H., \& Torquato, S. 2009, Proceedings of the National
  Academy of Sciences, 106, 17634

\bibitem[{{Kim} {et~al.}(2007){Kim}, {Bolton}, {Viel}, {Haehnelt}, \&
  {Carswell}}]{kim+07}
{Kim}, T., {Bolton}, J.~S., {Viel}, M., {Haehnelt}, M.~G., \& {Carswell}, R.~F.
  2007, \mnras, 382, 1657

\bibitem[{{Lai} {et~al.}(2006){Lai}, {Lidz}, {Hernquist}, \&
  {Zaldarriaga}}]{lai+06}
{Lai}, K., {Lidz}, A., {Hernquist}, L., \& {Zaldarriaga}, M. 2006, \apj, 644,
  61

\bibitem[{{Lidz} {et~al.}(2009){Lidz}, {Faucher-Giguere}, {Dall'Aglio},
  {McQuinn}, {Fechner}, {Zaldarriaga}, {Hernquist}, \& {Dutta}}]{lidz+09}
{Lidz}, A., {Faucher-Giguere}, C., {Dall'Aglio}, A., {McQuinn}, M., {Fechner},
  C., {Zaldarriaga}, M., {Hernquist}, L., \& {Dutta}, S. 2009, ArXiv e-prints

\bibitem[{{Lidz} {et~al.}(2006){Lidz}, {Heitmann}, {Hui}, {Habib}, {Rauch}, \&
  {Sargent}}]{lidz+06}
{Lidz}, A., {Heitmann}, K., {Hui}, L., {Habib}, S., {Rauch}, M., \& {Sargent},
  W.~L.~W. 2006, \apj, 638, 27

\bibitem[{{McDonald} {et~al.}(2001){McDonald}, {Miralda-Escud{\'e}}, {Rauch},
  {Sargent}, {Barlow}, \& {Cen}}]{mcd+01}
{McDonald}, P., {Miralda-Escud{\'e}}, J., {Rauch}, M., {Sargent}, W.~L.~W.,
  {Barlow}, T.~A., \& {Cen}, R. 2001, \apj, 562, 52

\bibitem[{{McDonald} {et~al.}(2000){McDonald}, {Miralda-Escud{\'e}}, {Rauch},
  {Sargent}, {Barlow}, {Cen}, \& {Ostriker}}]{mcd+00}
{McDonald}, P., {Miralda-Escud{\'e}}, J., {Rauch}, M., {Sargent}, W.~L.~W.,
  {Barlow}, T.~A., {Cen}, R., \& {Ostriker}, J.~P. 2000, \apj, 543, 1

\bibitem[{{McDonald} {et~al.}(2006){McDonald}, {Seljak}, {Burles}, {Schlegel},
  {Weinberg}, {Cen}, {Shih}, {Schaye}, {Schneider}, {Bahcall}, {Briggs},
  {Brinkmann}, {Brunner}, {Fukugita}, {Gunn}, {Ivezi{\'c}}, {Kent}, {Lupton},
  \& {Vanden Berk}}]{mcd+06}
{McDonald}, P., {Seljak}, U., {Burles}, S., {Schlegel}, D.~J., {Weinberg},
  D.~H., {Cen}, R., {Shih}, D., {Schaye}, J., 
  {et al.} 2006, \apjs, 163, 80

\bibitem[{{McDonald} {et~al.}(2005){McDonald}, {Seljak}, {Cen}, {Shih},
  {Weinberg}, {Burles}, {Schneider}, {Schlegel}, {Bahcall}, {Briggs},
  {Brinkmann}, {Fukugita}, {Ivezi{\'c}}, {Kent}, \& {Vanden Berk}}]{mcd+05}
{McDonald}, P., {Seljak}, U., {Cen}, R., {Shih}, D., {Weinberg}, D.~H.,
  {Burles}, S., {Schneider}, D.~P., {Schlegel}, D.~J., 
  {et al.} 2005, \apj, 635, 761

\bibitem[{{McQuinn} {et~al.}(2009){McQuinn}, {Lidz}, {Zaldarriaga},
  {Hernquist}, {Hopkins}, {Dutta}, \& {Faucher-Gigu{\`e}re}}]{mcquinn+09}
{McQuinn}, M., {Lidz}, A., {Zaldarriaga}, M., {Hernquist}, L., {Hopkins},
  P.~F., {Dutta}, S., \& {Faucher-Gigu{\`e}re}, C. 2009, \apj, 694, 842

\bibitem[{{Meiksin} \& {White}(2004)}]{meik+white04}
{Meiksin}, A. \& {White}, M. 2004, \mnras, 350, 1107

\bibitem[{{Miralda-Escud{\'e}} {et~al.}(1996){Miralda-Escud{\'e}}, {Cen},
  {Ostriker}, \& {Rauch}}]{mirald+96}
{Miralda-Escud{\'e}}, J., {Cen}, R., {Ostriker}, J.~P., \& {Rauch}, M. 1996,
  \apj, 471, 582

\bibitem[{{Schaye} {et~al.}(2000){Schaye}, {Theuns}, {Rauch}, {Efstathiou}, \&
  {Sargent}}]{schaye+00}
{Schaye}, J., {Theuns}, T., {Rauch}, M., {Efstathiou}, G., \& {Sargent},
  W.~L.~W. 2000, \mnras, 318, 817

\bibitem[{{Schneider} {et~al.}(2010){Schneider}, {Richards}, {Hall}, {Strauss},
  {Anderson}, {Boroson}, {Ross}, {Shen}, {Brandt}, {Fan}, {Inada}, {Jester},
  {Knapp}, {Krawczyk}, {Thakar}, {Vanden Berk}, {Voges}, {Yanny}, {York},
  {Bahcall}, {Bizyaev}, {Blanton}, {Brewington}, {Brinkmann}, {Eisenstein},
  {Frieman}, {Fukugita}, {Gray}, {Gunn}, {Hibon}, {Ivezi{\'c}}, {Kent}, {Kron},
  {Lee}, {Lupton}, {Malanushenko}, {Malanushenko}, {Oravetz}, {Pan}, {Pier},
  {Price}, {Saxe}, {Schlegel}, {Simmons}, {Snedden}, {SubbaRao}, {Szalay}, \&
  {Weinberg}}]{schneider+10}
{Schneider}, D.~P., {Richards}, G.~T., {Hall}, P.~B., {Strauss}, M.~A.,
  {Anderson}, S.~F., {Boroson}, T.~A., {Ross}, N.~P., {Shen}, Y., 
  {et al.} 2010, \aj, 139, 2360

\bibitem[{{Shull} {et~al.}(2010){Shull}, {France}, {Danforth}, {Smith}, \&
  {Tumlinson}}]{schull+10}
{Shull}, J.~M., {France}, K., {Danforth}, C.~W., {Smith}, B., \& {Tumlinson},
  J. 2010, \apj, 722, 1312

\bibitem[{{Slosar} {et~al.}(2009){Slosar}, {Ho}, {White}, \&
  {Louis}}]{slosar+09}
{Slosar}, A., {Ho}, S., {White}, M., \& {Louis}, T. 2009, Journal of Cosmology
  and Astro-Particle Physics, 10, 19

\bibitem[{{Stoughton} {et~al.}(2002){Stoughton}, {Lupton}, {Bernardi},
  {Blanton}, {Burles}, {Castander}, {Connolly}, {Eisenstein}, {Frieman},
  {Hennessy}, {Hindsley}, {Ivezi{\'c}}, {Kent}, {Kunszt}, {Lee}, {Meiksin},
  {Munn}, {Newberg}, {Nichol}, {Nicinski}, {Pier}, {Richards}, {Richmond},
  {Schlegel}, {Smith}, {Strauss}, {SubbaRao}, {Szalay}, {Thakar}, {Tucker},
  {Vanden Berk}, {Yanny}, {Adelman}, {Anderson}, {Anderson}, {Annis},
  {Bahcall}, {Bakken}, {Bartelmann}, {Bastian}, {Bauer}, {Berman},
  {B{\"o}hringer}, {Boroski}, {Bracker}, {Briegel}, {Briggs}, {Brinkmann},
  {Brunner}, {Carey}, {Carr}, {Chen}, {Christian}, {Colestock}, {Crocker},
  {Csabai}, {Czarapata}, {Dalcanton}, {Davidsen}, {Davis}, {Dehnen},
  {Dodelson}, {Doi}, {Dombeck}, {Donahue}, {Ellman}, {Elms}, {Evans}, {Eyer},
  {Fan}, {Federwitz}, {Friedman}, {Fukugita}, {Gal}, {Gillespie}, {Glazebrook},
  {Gray}, {Grebel}, {Greenawalt}, {Greene}, {Gunn}, {de Haas}, {Haiman},
  {Haldeman}, {Hall}, {Hamabe}, {Hansen}, {Harris}, {Harris}, {Harvanek},
  {Hawley}, {Hayes}, {Heckman}, {Helmi}, {Henden}, {Hogan}, {Hogg}, {Holmgren},
  {Holtzman}, {Huang}, {Hull}, {Ichikawa}, {Ichikawa}, {Johnston}, {Kauffmann},
  {Kim}, {Kimball}, {Kinney}, {Klaene}, {Kleinman}, {Klypin}, {Knapp},
  {Korienek}, {Krolik}, {Kron}, {Krzesi{\'n}ski}, {Lamb}, {Leger},
  {Limmongkol}, {Lindenmeyer}, {Long}, {Loomis}, {Loveday}, {MacKinnon},
  {Mannery}, {Mantsch}, {Margon}, {McGehee}, {McKay}, {McLean}, {Menou},
  {Merelli}, {Mo}, {Monet}, {Nakamura}, {Narayanan}, {Nash}, {Neilsen},
  {Newman}, {Nitta}, {Odenkirchen}, {Okada}, {Okamura}, {Ostriker}, {Owen},
  {Pauls}, {Peoples}, {Peterson}, {Petravick}, {Pope}, {Pordes}, {Postman},
  {Prosapio}, {Quinn}, {Rechenmacher}, {Rivetta}, {Rix}, {Rockosi}, {Rosner},
  {Ruthmansdorfer}, {Sandford}, {Schneider}, {Scranton}, {Sekiguchi}, {Sergey},
  {Sheth}, {Shimasaku}, {Smee}, {Snedden}, {Stebbins}, {Stubbs}, {Szapudi},
  {Szkody}, {Szokoly}, {Tabachnik}, {Tsvetanov}, {Uomoto}, {Vogeley}, {Voges},
  {Waddell}, {Walterbos}, {Wang}, {Watanabe}, {Weinberg}, {White}, {White},
  {Wilhite}, {Wolfe}, {Yasuda}, {York}, {Zehavi}, \& {Zheng}}]{stoughton+02}
{Stoughton}, C., {Lupton}, R.~H., {Bernardi}, M., {Blanton}, M.~R., {Burles},
  S., {Castander}, F.~J., {Connolly}, A.~J., {Eisenstein}, D.~J., 
  {et al.} 2002,
  \aj, 123, 485

\bibitem[{{Suzuki} {et~al.}(2005){Suzuki}, {Tytler}, {Kirkman}, {O'Meara}, \&
  {Lubin}}]{suz+05}
{Suzuki}, N., {Tytler}, D., {Kirkman}, D., {O'Meara}, J.~M., \& {Lubin}, D.
  2005, \apj, 618, 592

\bibitem[{{Theuns} {et~al.}(2002){Theuns}, {Bernardi}, {Frieman}, {Hewett},
  {Schaye}, {Sheth}, \& {Subbarao}}]{theuns+02}
{Theuns}, T., {Bernardi}, M., {Frieman}, J., {Hewett}, P., {Schaye}, J.,
  {Sheth}, R.~K., \& {Subbarao}, M. 2002, \apjl, 574, L111

\bibitem[{{Theuns} {et~al.}(1998){Theuns}, {Leonard}, {Efstathiou}, {Pearce},
  \& {Thomas}}]{theuns+98}
{Theuns}, T., {Leonard}, A., {Efstathiou}, G., {Pearce}, F.~R., \& {Thomas},
  P.~A. 1998, \mnras, 301, 478

\bibitem[{Torquato {et~al.}(1988)Torquato, Beasley, \& Chiew}]{torq+88}
Torquato, S., Beasley, J.~D., \& Chiew, Y.~C. 1988, The Journal of Chemical
  Physics, 88, 6540

\bibitem[{{Viel} {et~al.}(2009){Viel}, {Bolton}, \& {Haehnelt}}]{viel+09}
{Viel}, M., {Bolton}, J.~S., \& {Haehnelt}, M.~G. 2009, \mnras, 399, L39

\bibitem[{{White} {et~al.}(2010){White}, {Pope}, {Carlson}, {Heitmann},
  {Habib}, {Fasel}, {Daniel}, \& {Lukic}}]{white+10}
{White}, M., {Pope}, A., {Carlson}, J., {Heitmann}, K., {Habib}, S., {Fasel},
  P., {Daniel}, D., \& {Lukic}, Z. 2010, \apj, 713, 383

\bibitem[{{York} {et~al.}(2000){York}, {Adelman}, {Anderson}, {Anderson},
  {Annis}, {Bahcall}, {Bakken}, {Barkhouser}, {Bastian}, {Berman}, {Boroski},
  {Bracker}, {Briegel}, {Briggs}, {Brinkmann}, {Brunner}, {Burles}, {Carey},
  {Carr}, {Castander}, {Chen}, {Colestock}, {Connolly}, {Crocker}, {Csabai},
  {Czarapata}, {Davis}, {Doi}, {Dombeck}, {Eisenstein}, {Ellman}, {Elms},
  {Evans}, {Fan}, {Federwitz}, {Fiscelli}, {Friedman}, {Frieman}, {Fukugita},
  {Gillespie}, {Gunn}, {Gurbani}, {de Haas}, {Haldeman}, {Harris}, {Hayes},
  {Heckman}, {Hennessy}, {Hindsley}, {Holm}, {Holmgren}, {Huang}, {Hull},
  {Husby}, {Ichikawa}, {Ichikawa}, {Ivezi{\'c}}, {Kent}, {Kim}, {Kinney},
  {Klaene}, {Kleinman}, {Kleinman}, {Knapp}, {Korienek}, {Kron}, {Kunszt},
  {Lamb}, {Lee}, {Leger}, {Limmongkol}, {Lindenmeyer}, {Long}, {Loomis},
  {Loveday}, {Lucinio}, {Lupton}, {MacKinnon}, {Mannery}, {Mantsch}, {Margon},
  {McGehee}, {McKay}, {Meiksin}, {Merelli}, {Monet}, {Munn}, {Narayanan},
  {Nash}, {Neilsen}, {Neswold}, {Newberg}, {Nichol}, {Nicinski}, {Nonino},
  {Okada}, {Okamura}, {Ostriker}, {Owen}, {Pauls}, {Peoples}, {Peterson},
  {Petravick}, {Pier}, {Pope}, {Pordes}, {Prosapio}, {Rechenmacher}, {Quinn},
  {Richards}, {Richmond}, {Rivetta}, {Rockosi}, {Ruthmansdorfer}, {Sandford},
  {Schlegel}, {Schneider}, {Sekiguchi}, {Sergey}, {Shimasaku}, {Siegmund},
  {Smee}, {Smith}, {Snedden}, {Stone}, {Stoughton}, {Strauss}, {Stubbs},
  {SubbaRao}, {Szalay}, {Szapudi}, {Szokoly}, {Thakar}, {Tremonti}, {Tucker},
  {Uomoto}, {Vanden Berk}, {Vogeley}, {Waddell}, {Wang}, {Watanabe},
  {Weinberg}, {Yanny}, \& {Yasuda}}]{york+00}
{York}, D.~G., {Adelman}, J., {Anderson}, Jr., J.~E., {Anderson}, S.~F.,
  {Annis}, J., {Bahcall}, N.~A., {Bakken}, J.~A., {Barkhouser}, R.,  {et al.} 2000, \aj, 120, 1579

\bibitem[{{Zaldarriaga} {et~al.}(2003){Zaldarriaga}, {Scoccimarro}, \&
  {Hui}}]{zald+03}
{Zaldarriaga}, M., {Scoccimarro}, R., \& {Hui}, L. 2003, \apj, 590, 1

\end{thebibliography}
\end{document}